\documentclass[usenatbib]{mn2e}

\usepackage{apjfonts,amsfonts,verbatim}
\usepackage{bm}
\usepackage{mathtools}
\usepackage{xcolor}
\usepackage{natbib}
\usepackage{lscape}
\usepackage{afterpage}  
\usepackage{booktabs}
\usepackage{tabularx}
\usepackage{adjustbox}

\usepackage{threeparttable}

\usepackage[T1]{fontenc}

\usepackage{graphicx}	
\usepackage{amsmath}	
\usepackage{amssymb}	

\newcommand{\lya}{Ly$\alpha$}

\newcommand{\hb}{H$\beta$}
\newcommand{\oiii}{[\rm O\,{\textsc {iii}}]}

\newcommand{\HI}{\rm H\,{\textsc {i}}}

\newcommand{\kms}{\,\ifmmode{\mathrm{km}\,\mathrm{s}^{-1}}\else km\,s${}^{-1}$\fi}

\title[Gas Kinematics in LAB2]{Where Outflows Meet Inflows: Gas Kinematics in SSA22 Lyman-$\alpha$ Blob 2 Decoded by Advanced Radiative Transfer Modelling}

\author[Li et al.]{
Zhihui Li,$^{1}$\thanks{E-mail: zhihui@caltech.edu}
Charles C. Steidel,$^{1}$
Max Gronke$^{2,3}$\thanks{Hubble fellow}, Yuguang Chen$^{1}$ and Yuichi Matsuda$^{3,4}$
\\
$^{1}$Cahill Center for Astrophysics, California Institute of Technology, MC 249-17, 1200 East California Boulevard, Pasadena, CA 91125, USA\\
$^{2}$Max-Planck Institute for Astrophysics, Karl-Schwarzschild-Str. 1, D-85741 Garching, Germany\\
$^{3}$Department of Physics \& Astronomy, Johns Hopkins University, Baltimore, MD 21218, USA\\
$^{4}$National Astronomical Observatory of Japan, 2-21-1 Osawa, Mitaka, Tokyo 181-8588, Japan\\
$^{5}$Department of Astronomy, School of Science, The Graduate University for Advanced Studies, SOKENDAI, Osawa, Mitaka, Tokyo 181-8588, Japan
}

\date{}

\begin{document}
\label{firstpage}
\pagerange{\pageref{firstpage}--\pageref{lastpage}}
\maketitle

\begin{abstract}
We present new spectroscopic observations of Lyman-$\alpha$ (\lya) Blob 2 ($z \sim$ 3.1). We observed extended \lya\ emission in three distinct regions, where the highest \lya\ surface brightness (SB) center is far away from the known continuum sources. We searched through the MOSFIRE slits that cover the high \lya\ SB regions, but were unable to detect any significant nebular emission near the highest SB center. We further mapped the flux ratio of the blue peak to the red peak and found it is anti-correlated with \lya\ SB with a power-law index of $\sim$ -- 0.4. We used radiative transfer models with both multiphase, clumpy and shell geometries and successfully reproduced the diverse \lya\ morphologies. We found that most spectra suggest outflow-dominated kinematics, while 4/15 spectra imply inflows. A significant correlation exists between parameter pairs, and the multiphase, clumpy model may alleviate previously reported discrepancies. We also modeled \lya\ spectra at different positions simultaneously and found that the variation of the inferred clump outflow velocities can be approximately explained by line-of-sight projection effects. Our results support the `central powering + scattering' scenario, i.e. the \lya\ photons are generated by a central powering source and then scatter with outflowing, multiphase \HI\ gas while propagating outwards. The infalling of cool gas near the blob outskirts shapes the observed blue-dominated \lya\ profiles, but its energy contribution to the total \lya\ luminosity is less than 10\%, i.e. minor compared to the photo-ionization by star-forming galaxies and/or AGNs.

\end{abstract}


\section{Introduction}
Lyman-$\alpha$ blobs (LABs) -- spatially extended (projected sizes $\gtrsim$\,100 kpc) gaseous nebulae with tremendous \lya\ luminosities ($L_{\rm Ly\alpha} \sim 10^{43-44}$\,erg\,$\rm s^{-1}$) seen at high redshifts ($z \gtrsim$ 2) -- are among the most mysterious and intriguing objects in the universe. Thus far, numerous LABs have been discovered via narrow-band imaging or in galaxy surveys (e.g. \citealt{Francis96, Fynbo99, Keel99, Steidel2000, Matsuda04, Matsuda11, Dey05, Saito06, SmithJarvis07, Hennawi09, Ouchi09, Prescott09, Prescott12, Erb11, Cai17}), yet their physical origin remains obscure. It is found that LABs are preferentially located in overdense protocluster regions, which are expected to be the progenitors of the massive galaxy clusters observed today (e.g. \citealt{Steidel98, Prescott08, Yang09, Yang10, Hine16}). Therefore, the study of LABs may illuminate the formation and evolution of massive galaxies and the mechanisms of associated feedback events.

Up to now, many hypotheses about the powering mechanisms of LABs have been proposed, including: (1) Photo-ionization by central energetic sources (starburst galaxies or AGNs, see e.g. \citealt{Haiman01, Cantalupo05, Cantalupo14}). This scenario gained credence from infrared and submillimeter observations that discovered luminous galaxies and AGNs associated with some LABs \citep{Chapman01, Dey05, Geach05, Geach07, Geach09, Colbert06, Webb09}. (2) Starburst-induced, shock-powered galactic-wide outflows (`superwinds', see e.g. \citealt{Heckman90, Taniguchi00, Taniguchi01, Mori04}). This scenario has been corroborated by the observed redward asymmetric \lya\ profiles \citep{Dawson02, Ajiki02} and shell-like or bubble-like structures \citep{Matsuda04, Wilman05}. (3) Gravitational cooling radiation from accretion of cold gas streams onto protogalaxies (e.g. \citealt{Haiman00, Fardal01, Furlanetto05, Dijkstra06a, Dijkstra06b, Goerdt10, Faucher10, Rosdahl12}). This scenario may be preferred for LABs that appear to lack powerful sources despite deep multi-wavelength observations and exhibit blueward asymmetric \lya\ profiles \citep{Nilsson06, SmithJarvis07, Saito08, Smith08, Daddi20}. (4) Resonant scattering of \lya\ photons (e.g. \citealt{Dijkstra08, Steidel10, Steidel11}). As resonant scattering imposes polarization, recent polarimetric observations and simulations have provided evidence of scattering within LABs, albeit with remaining uncertainties \citep{Dijkstra09, Hayes11, Beck16, Trebitsch16, Eide2018, Kim20}.

To further distinguish these different powering mechanisms, it is beneficial to study spatially-resolved \lya\ spectra, which are made possible by the outstanding capabilities of recently commissioned integral field unit (IFU) spectrographs, such as KCWI (Keck Cosmic Web Imager, \citealt{KCWI18}) and MUSE (Multi Unit Spectroscopic Explorer, \citealt{Bacon14}). These instruments have revolutionized the study of extended \lya\ nebulae by adding an additional spatial dimension with unprecedented sensitivity at rest-frame UV wavelengths (see e.g. \citealt{Wisotzki16,Leclercq17,Leclercq20}).

In this paper, we present new KCWI observations of one of the giant LABs discovered in the overdense proto-cluster region SSA22 at $z \sim$ 3.1, SSA22-Blob2 (LAB2, \citealt{Steidel98, Steidel2000}). LAB2 has an immense \lya\ luminosity of $\sim$\,10$^{44}$\,erg\,s$^{-1}$ and a spatial extent of $\sim$\,100\,kpc. Ever since its discovery, LAB2 has become the target of many follow-up observations at multiple wavelengths, including X-ray \citep{Basu04, Geach09, Lehmer09cat, Lehmer09AGN}, optical \citep{Wilman05, Martin14}, infrared (IR, \citealt{Geach07, Webb09}), and submillimeter (submm, \citealt{Chapman01, Hine16b}). A Lyman-break galaxy (LBG), M14 \citep{Steidel2000}, and an X-ray source \citep{Basu04} have been identified within LAB2. 

To analyse the spatially-resolved \lya\ profiles obtained by KCWI, we carried out Monte-Carlo radiative transfer (MCRT) modelling assuming a multiphase, clumpy \HI\ gas model. As a presumably realistic description of the \HI\ gas in the interstellar medium (ISM) and the circumgalactic medium (CGM), the multiphase, clumpy model has been explored by many theoretical studies (e.g. \citealt{Neufeld91, Hansen06, Dijkstra12, Laursen13, Gronke16_model}). Observationally, \citet{Li21} made the first successful attempt to systemically fit the spatially-resolved \lya\ profiles with the multiphase, clumpy model. The present work is a direct follow-up of \citet{Li21}, exploring a different parameter space with new physical interpretation of the derived parameters.

The structure of this paper is as follows. In \S\ref{sec:data}, we describe our KCWI and MOSFIRE observations and data reduction procedures. In \S\ref{sec:kinematic}, we present the spatial distribution and spectral profiles of \lya\ emission. In \S\ref{sec:oiii}, we present the non-detection of nebular emission within LAB2 with MOSFIRE. In \S\ref{sec:RT}, we detail the methodology and present the results of radiative transfer modelling using the multiphase, clumpy model. In \S\ref{sec:previous}, we summarize previous studies of LAB2 and compare with this work. In \S\ref{sec:conclusion}, we summarize and conclude. Throughout this paper we adopt a flat $\Lambda$CDM cosmology with $\Omega_{\rm m}$ = 0.315, $\Omega_{\Lambda}$ = 0.685, and $H_{0}$ = 67.4 km s$^{-1}$ Mpc$^{-1}$ \citep{Planck18}. We use the following vacuum wavelengths: 1215.67\,\AA\ for \lya, 4862.683\,\AA\ for H$\beta$, and 4960.295/5008.240\,\AA\ for \oiii\ from the Atomic Line List v2.04\footnote{http://www.pa.uky.edu/$\sim$peter/atomic/index.html}.

\section{Observations and Data Reduction}\label{sec:data}

\subsection{KCWI Observations}\label{sec:KCWIdata}
The KCWI observations of LAB2 were carried out on the night of 2019 September 27, with a seeing of $\sim 0.4 - 0.5\arcsec$ full width at half maximum (FWHM). We used the KCWI large slicer, which provides a contiguous field-of-view of 20.4$\arcsec$ (slice length) $\times$ 33$\arcsec$ (24 $\times$ 1.35\arcsec\ slice width). With the BM VPH grating set up for $\lambda_{\rm c} = 4800$\,\AA, the wavelength coverage is $\sim 4260 - 5330$\,\AA, with spectral resolution $R\simeq 1800-2200$. The data were obtained as 8 individual 1200\,s exposures, with small telescope offsets in the direction perpendicular to slices applied between each, in an effort to recover some spatial resolution given the relatively large slice width. The total on-source exposure time was 2.7 hours, and the SB detection limit (1$\sigma$) is about 8 $\times$ 10$^{-20}$ erg\,s$^{-1}$\,cm$^{-2}$\,arcsec$^{-2}$ per seeing element (1 arcsec$^{2}$) using an unresolved emission line near the \lya\ emission wavelength at $z \sim 3.09$.

Individual exposures were reduced using the KCWI Data Reduction Pipeline (DRP)\footnote{https://github.com/Keck-DataReductionPipelines/KcwiDRP}, which includes wavelength calibration, atmospheric refraction correction, background subtraction, and flux calibration. The sky subtraction was conducted using both the DRP, which constructs a b-spline sky model, and a custom median filtering procedure to remove the low-order scattered light. Both the continuum and line emission sources were masked. For more details, we refer the readers to \citet{Chen21}, in which the data reduction procedures were adopted directly in this paper. The individual datacubes were then spatially re-sampled onto a uniform astrometric grid with 0.3\arcsec\ by 0.3\arcsec\ spaxels, with a sampling of 0.5\,\AA\,pix$^{-1}$ (4.75 pixels per spectral resolution element) along the wavelength axis, using a variant of the `drizzle' algorithm (with a drizzle factor of 0.9) in the \texttt{MONTAGE}\footnote{http://montage.ipac.caltech.edu} package. The re-sampled cubes were then combined into a final stacked cube by averaging with exposure time weighting. Owing to the coarser spatial sampling in the long dimension of the spatial cube, the PSF in the final datacube is elongated along the N-S direction, with FWHM $\simeq 0.90\arcsec \times 1.08\arcsec$ (X-direction and Y-direction, respectively) measured from the most compact object in the field. We also conducted astrometry by cross-correlating the pseudo-white-light images from the KCWI datacubes to the existing wide-field astrometry-corrected images (e.g. the \emph{HST} WFC3-IR F160W image in our Figure \ref{fig:Lya_images}). This procedure is done for both individual exposures and the full stack of exposures, thus providing both relative and absolute astrometric information. An alignment with the \emph{HST} image was done after drizzling and combining the individual KCWI exposures.
 
The resampled final datacube covers a scientifically useful solid angle of 22.6$\arcsec \times 33.6\arcsec$ on the sky. A variance image with the same dimensions was created by propagating errors based on a noise model throughout the data reduction. 

\subsection{MOSFIRE Observations}\label{sec:MOSFIREdata}

We observed selected regions of LAB2, chosen to include the highest \lya\ surface brightness areas, using MOSFIRE \citep{McLean10, McLean12,Steidel14} on the Keck I telescope. Spectra in the near-IR $K$ band (1.967 -- 2.393\,${\mu}$m) were obtained using two slitmasks with the same sky PA, which cover two parallel regions of width 1\arcsec\ separated by 0.25\arcsec\ on the sky, as shown in Figure \ref{fig:Lya_images} and summarised in Table \ref{tab:mosfire_obs}. Slit 1 passes through the region with the highest \lya\ surface brightness (labeled 11 in Fig.~\ref{fig:Lya_images}), while Slit 2 abuts that region immediately to the south, and also includes a second high SB region (labeled 13) to the southeast. 

The MOSFIRE observations of Slit 1 and Slit 2 were obtained on two separate observing runs (on 2020 November 27 and 2019 June 15, respectively), both under clear conditions with sub-arcsec seeing. With 1.0\arcsec\ slits in the $K$ band, MOSFIRE achieves a spectral resolving power $R \sim 2660$. The data were taken with the MOSFIRE ``masknod" mode, using two telescope positions separated by 30 -- 40\arcsec\ along the slit direction, with individual exposures of 180\,s between nods. The total integration times were 2.0 hours for Slit 1 and 2.5 hours for Slit 2. The data for each observation sequence were reduced using the MOSFIRE data reduction pipeline\footnote{https://github.com/Keck-DataReductionPipelines/MosfireDRP} to produce two-dimensional, rectified, background-subtracted vacuum wavelength calibrated spectrograms (see \citealt{Steidel14} for details). Observations obtained on different observing nights using the same slitmask were reduced independently; the 2-D spectrograms were shifted into the heliocentric rest frame and combined with inverse variance weighting using tasks in the \texttt{MOSPEC} \citep{Strom17} analysis package.

\begin{table}
    \centering
    \scriptsize \caption{MOSFIRE $K$-band observations of LAB2.}
    \label{tab:mosfire_obs}
    \begin{tabular}{cccccccc}
    \hline\hline
    Name & Width & $R$ & PA & Exp & Seeing & Date of Obs & Nod  \\ 
     (1)  & (2) & (3) & (4) & (5) & (6) & (7) & (8) \\
    \hline
    Slit 1 & 1.0 & 2660 & $-54.0$ & 2.0 & 0.71 & 2020 Nov 27 & $\pm 20.0$\\
    Slit 2 & 1.0 & 2660 & $-54.0$ &  2.5 & 0.45 & 2019 Jun 15 & $\pm 15.0$\\
    \hline\hline
    \end{tabular}
    \begin{tablenotes}
    \item \textbf{Notes.} The details of the MOSFIRE $K$-band observations of LAB2. The columns are: (1) slit name; (2) slit width ($\arcsec$); (3) resolving power ($\lambda/\Delta\lambda$); (4) slit PA (degrees E of N); (5) exposure time in hours; (6) seeing FWHM ($\arcsec$); (7) UT date of observation; (8) nod amplitude between A and B positions ($\arcsec$).
    \end{tablenotes}
\end{table}

\begin{table}
    \centering
    \scriptsize \caption{Continuum sources identified in LAB2.}
	\label{tab:sources}
	\renewcommand{\arraystretch}{1.2}
	\begin{tabular}{cccccc} 
		\hline \hline
		Name & RA (J2000) & Dec (J2000) & $z_{\rm sys}$ & Type & Refs.\\
		\hline  
		M14$^{\rm a}$ & 22:17:39.09 & +00:13:29.8 & 3.091 &\lya\ & (1)(2)\\
		LAB2-a & 22:17:39.3 & +00:13:22.0 & ... & IR & (3)\\
		LAB2-b & 22:17:39.1 & +00:13:30.7 & ... & IR & (3)\\
		LAB2-ALMA & 22:17:39.079 & +00:13:30.85 & ... & Submm & (4)\\
		LAB2-X-ray & 22:17:39.08 & +00:13:30.7 & ... &X-ray & (5)\\
		\hline \hline
	\end{tabular}
	\begin{tablenotes}
    \item $^{\rm a}$Originally defined in \citet{Steidel2000}. \\
    \item \textbf{Notes.} Properties of the continuum sources identified in LAB2. The columns are: (1) name of the source; (2) right ascension; (3) declination; (4) systemic redshift; (5) type of observation; (6) references. \\
    \textbf{References.} (1) \citet{Steidel03}; (2) \citet{Nestor13}; (3) \citet{Webb09}; (4) \citet{Ao17}; (5) \citet{Lehmer09cat}.
    \end{tablenotes}
\end{table}

\section{\lya\ Emission in LAB2}\label{sec:kinematic}

\subsection{Spatial Distribution of Ly$\alpha$ Emission}\label{sec:spatial}

\begin{figure*}
\centering
\includegraphics[width=0.498\textwidth]{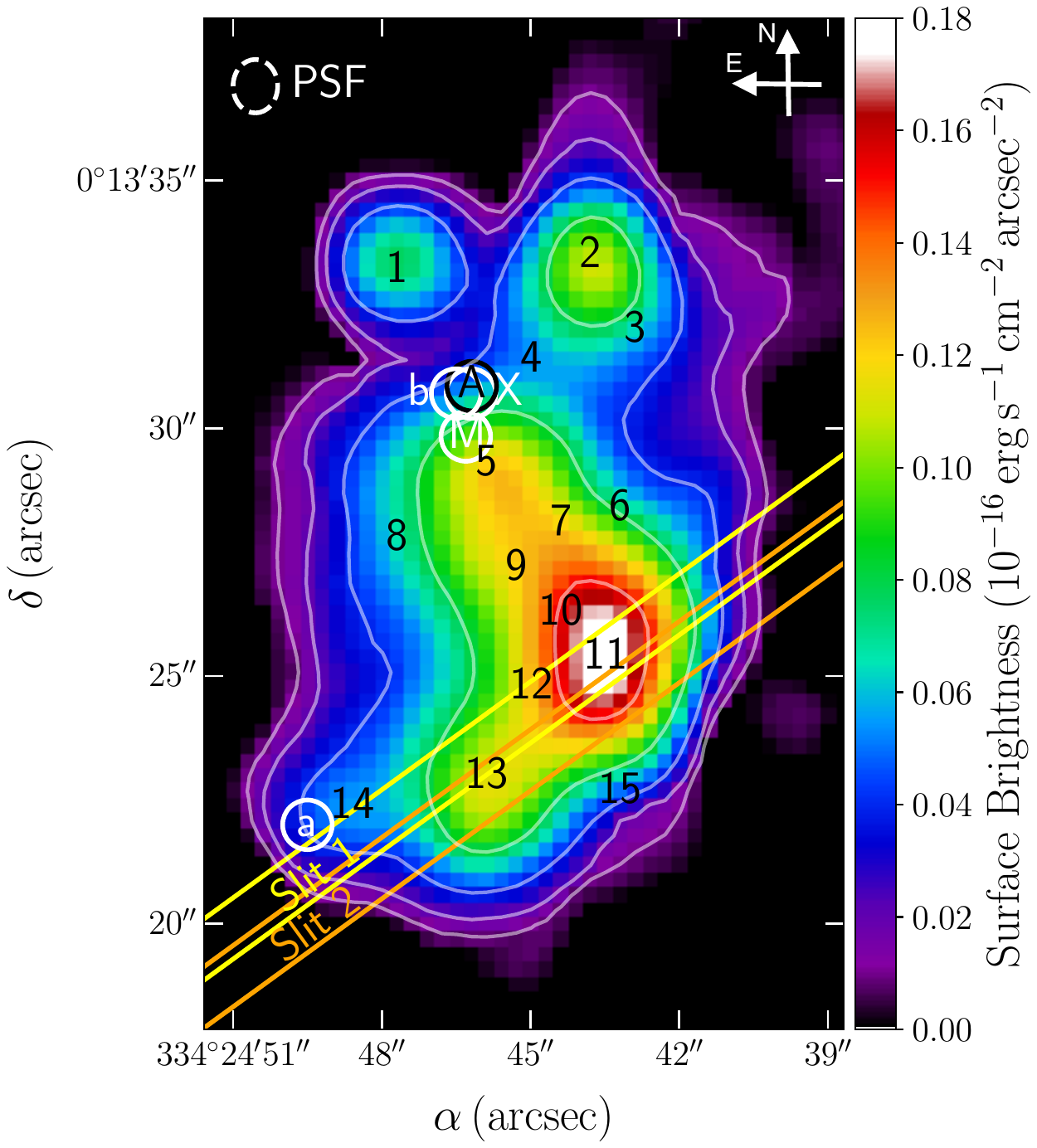}
\includegraphics[width=0.414\textwidth]{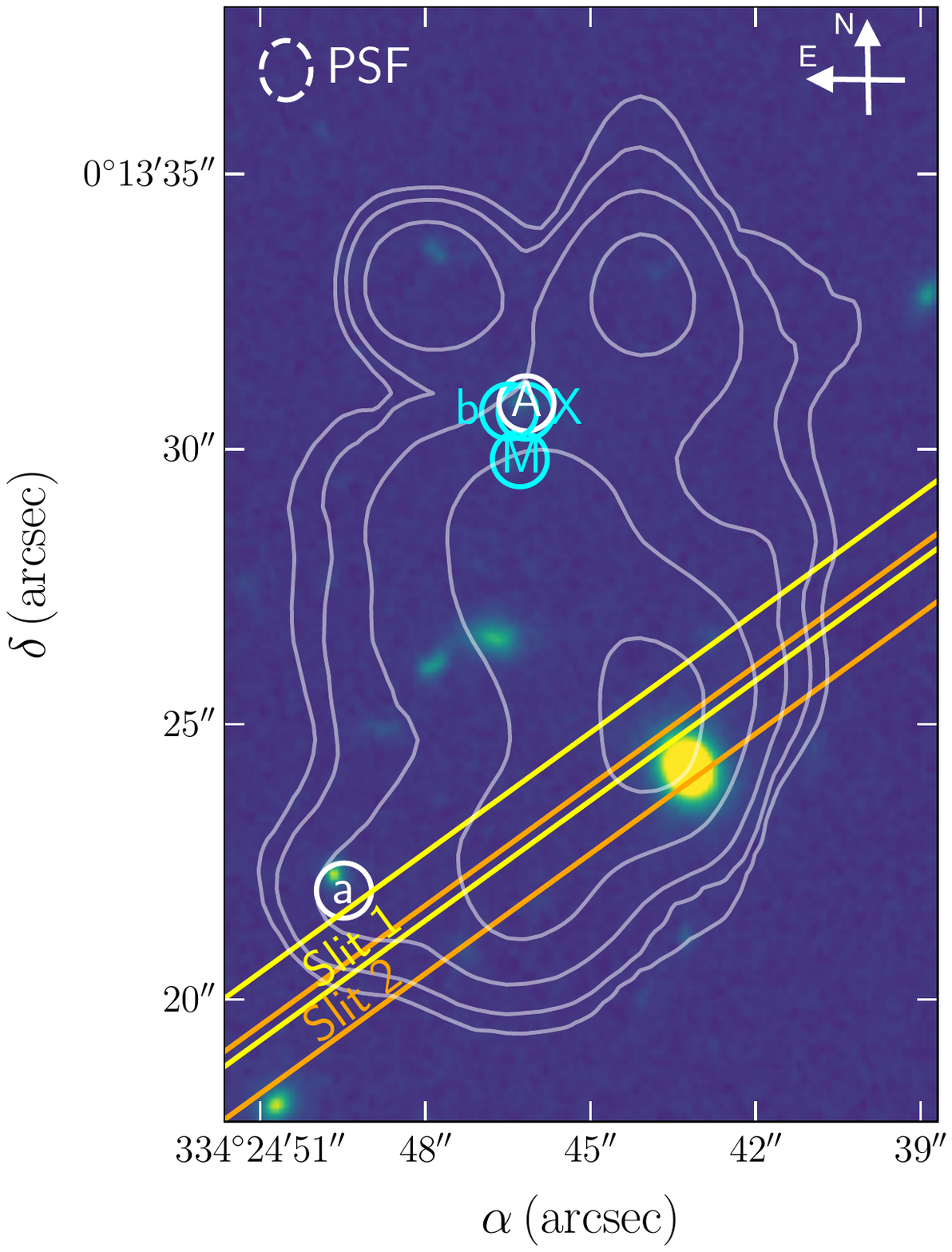}\\
    \caption{\lya\ and continuum images of LAB2. \emph{Left}: The narrow band \lya\ image, obtained by collapsing the original KCWI datacube over 4949 -- 5009\,\AA, which encloses the \lya\ line (see \S\ref{sec:spatial}). The UV continuum near the wavelength of \lya\ has been subtracted. The positions of the MOSFIRE slits are delineated by parallel yellow and orange lines (see \S\ref{sec:MOSFIREdata}), and the numbers (1-15, in black) indicate the positions of spectra that we sample for radiative transfer modelling in \S\ref{sec:RT}. \emph{Right}: The \emph{HST} WFC3-IR F160W rest-frame optical continuum image. The positions of a Lyman-break galaxy (M14, labeled as `M'), an X-ray source (labeled as `X') with IR (`b') and submm (`A') counterparts, and an IR source (`a') have been marked on each image (see \S\ref{sec:profiles} and Table \ref{tab:sources}). The \lya\ isophotes with levels of SB$_{\rm Ly\alpha}$ = [150,\,80,\,40,\,20,\,10]$\,\times\,10^{-19}$\,erg\,s$^{-1}$\,cm$^{-2}$\,arcsec$^{-2}$ have also been overlaid, and the dashed white ellipse indicates the PSF in the final datacube with FWHM $\simeq 0.90\arcsec \times 1.08\arcsec$ (X and Y-direction, respectively). Both images have been registered to the same world-coordinate system.
    \label{fig:Lya_images}}
\end{figure*}

\begin{figure*}
\centering
\includegraphics[width=0.98\textwidth]{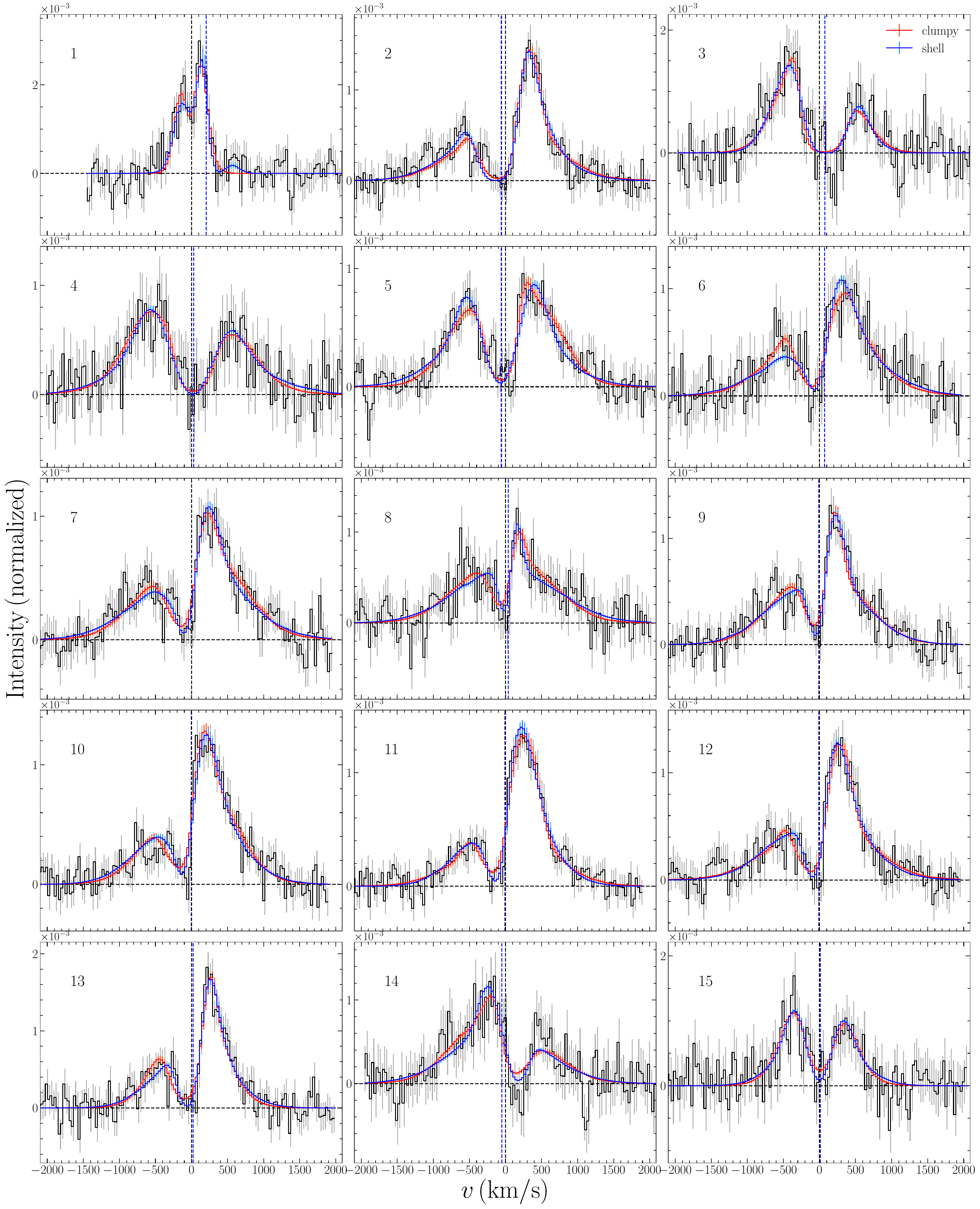}
    \caption{Fifteen representative continuum-subtracted, spatially-resolved and normalised \lya\ profiles (black, with grey 1-$\sigma$ error bars) from the high SB regions in LAB2. The spectrum number of each spectrum has been marked on the SB map in Figure \ref{fig:Lya_images}. All the spectra have been smoothed by a 3 pixel $\times$ 3 pixel boxcar (0.9$\arcsec$) spatially and Gaussian smoothed ($\sigma$ = 0.5\,\AA) in the wavelength dimension. As we will detail in \S\ref{sec:correlations}, the multiphase, clumpy model best-fits (red, with orange 1-$\sigma$ Poisson errors) and the shell model best-fits (blue, with cyan 1-$\sigma$ Poisson errors) are both shown in each subpanel. The observed \lya\ spectra have also been shifted by --$\Delta v_{\rm clumpy}$ to their local systemic redshifts (as determined by the best-fits), and the shell model best-fits are shifted correspondingly as well for direct comparison. For each subpanel, the $x$-axis is the velocity (in km\,s$^{-1}$) with respect to the local systemic redshift, and the $y$-axis is the normalised line flux. For visual reference, the horizontal and vertical black dashed lines in each subpanel indicate zero flux level and zero velocity with respect to the local systemic redshift, respectively. The vertical blue dashed lines indicate the initial guess for the systemic redshift ($z$ = 3.09 for spectrum 1 and $z$ = 3.098 for all other spectra).
    \label{fig:Lya_linemaps}}
\end{figure*}   

In preparation for our following \lya\ analyses, we first smoothed the KCWI datacube spatially with a 3 pixel $\times$ 3 pixel boxcar, and spectrally with a $\sigma$ = 0.5\,\AA\,Gaussian function. Such a smoothed datacube will be used for all the following \lya\ spectral analyses in this work. To provide an overview of the \lya\ surface brightness (SB) distribution in LAB2, we generated a \lya\ narrow-band image by summing all the \lya\ fluxes over the relevant wavelength range. Similar to \citet{Li21}, we followed the `matched filtering' procedures using \texttt{LSDCat} \citep{Herenz17}, specifically with three steps\footnote{Note that the \texttt{LSDCat} procedures are only applied in this section for generating the Ly$\alpha$ narrow-band image.}: (1) applying spatial filtering to the aforementioned smoothed KCWI datacube using a 2D Gaussian filter with FWHM = 0.9$\arcsec$ (the seeing point spread function (PSF) measured from a bright star in the SSA22 field); (2) applying a 1D Gaussian spectral filter with FWHM = 500\,km\,s$^{-1}$ (a conservative lower limit on the observed \lya\ line width estimated visually); (3) generating an S/N cube with the filtered datacube for thresholding.

We present the \lya\ narrow-band image of LAB2 in the left panel of Figure \ref{fig:Lya_images}. It is constructed by summing all the voxels of the continuum-subtracted filtered datacube with S/N $\geq$ 4 over 4949 -- 5009\,\AA\ ($\sim$ --2000 -- 1500\,km\,s$^{-1}$ at $z \sim$ 3.1), which should enclose all possible \lya\ emission. The level of the subtracted continuum is determined as the average of the two median UV flux densities bluewards (4849 -- 4949\,\AA) and redwards (5009 -- 5109\,\AA) of the \lya\ emission. In the right panel, we present the \emph{HST} WFC3-IR F160W rest-frame optical continuum image (Program 13844, PI: Bret Lehmer) of LAB2\footnote{The KCWI and \emph{HST} images have been registered to the same world-coordinate system using cross-correlation.} for comparison. The positions of previously identified sources are marked on each image as references. Thus far, multiple sources have been identified in LAB2: an LBG M14 \citep{Steidel2000}, an X-ray source \citep{Basu04, Lehmer09cat} with IR (LAB2-b, \citealt{Webb09}) and submm (LAB2-ALMA, \citealt{Ao17}) counterparts, and an IR source detected by \emph{Spitzer} IRAC (LAB2-a, \citealt{Webb09}). The detailed information for all the identified sources are presented in Table \ref{tab:sources}. The \lya\ isophotes (contours with the same SB) with levels of SB$_{\rm Ly\alpha}$ = [150,\,80,\,40,\,20,\,10]$\,\times\,10^{-19}$\,erg\,s$^{-1}$\,cm$^{-2}$\,arcsec$^{-2}$ have also been overlaid onto each image.

In Figure \ref{fig:Lya_images}, we see that the extended \lya\ emission can be separated mainly in three distinct regions -- the northeast (as indicated by a number `1'), the northwest (as indicated by a number `2'), and the south region. Among them, the south region has the largest extent. In this extended region there appears to be a high SB center (as indicated by a number `11'), yet neither M14 nor the X-ray source is close to this position -- instead, they are located in the northeast outskirts where the \lya\ SB is relatively low. In addition, the variation of the \lya\ SB with respect to the maximum value is highly direction-dependent. Two regions exhibiting monotonically declining \lya\ SB towards the northeast and southeast directions are evident, the former of which is more elongated.

\subsection{Profiles of \lya\ Emission \& Blue-to-Red Flux Ratio}\label{sec:profiles}

\begin{figure*}
\centering
\includegraphics[width=0.359\textwidth]{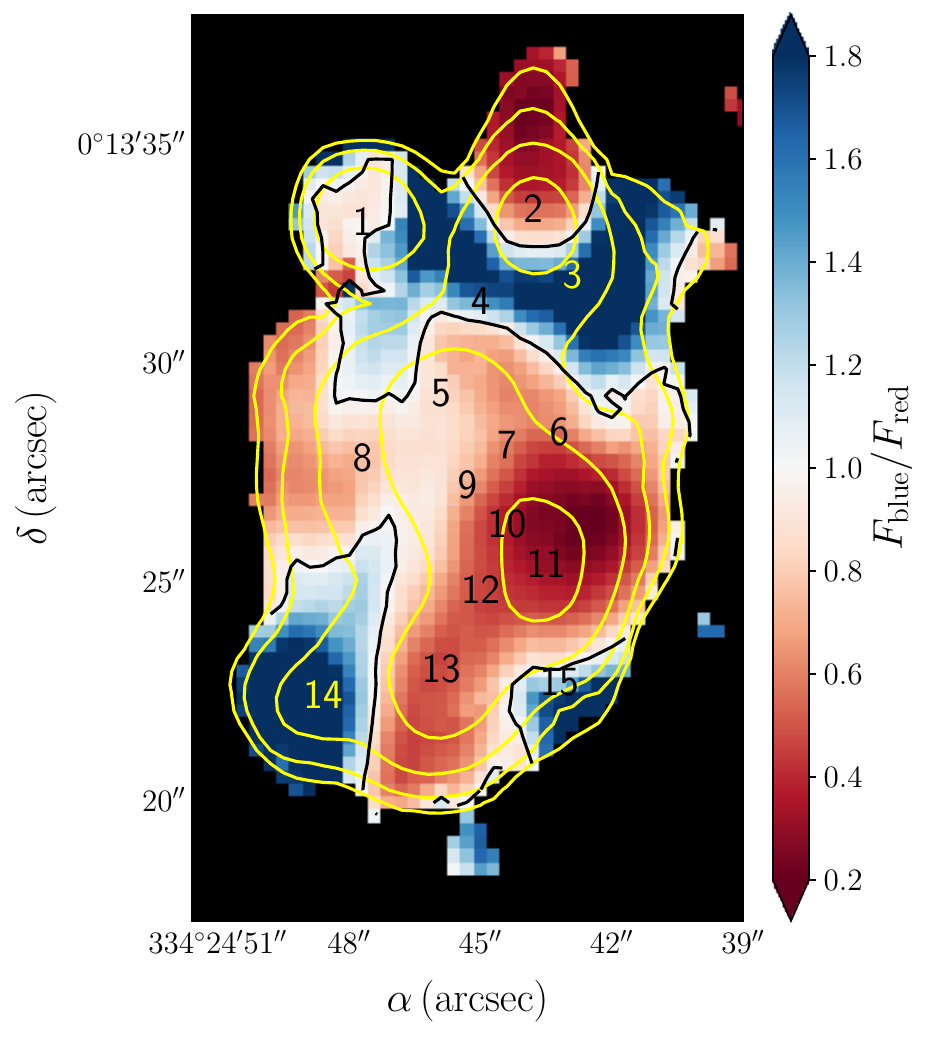}
\includegraphics[width=0.632\textwidth]{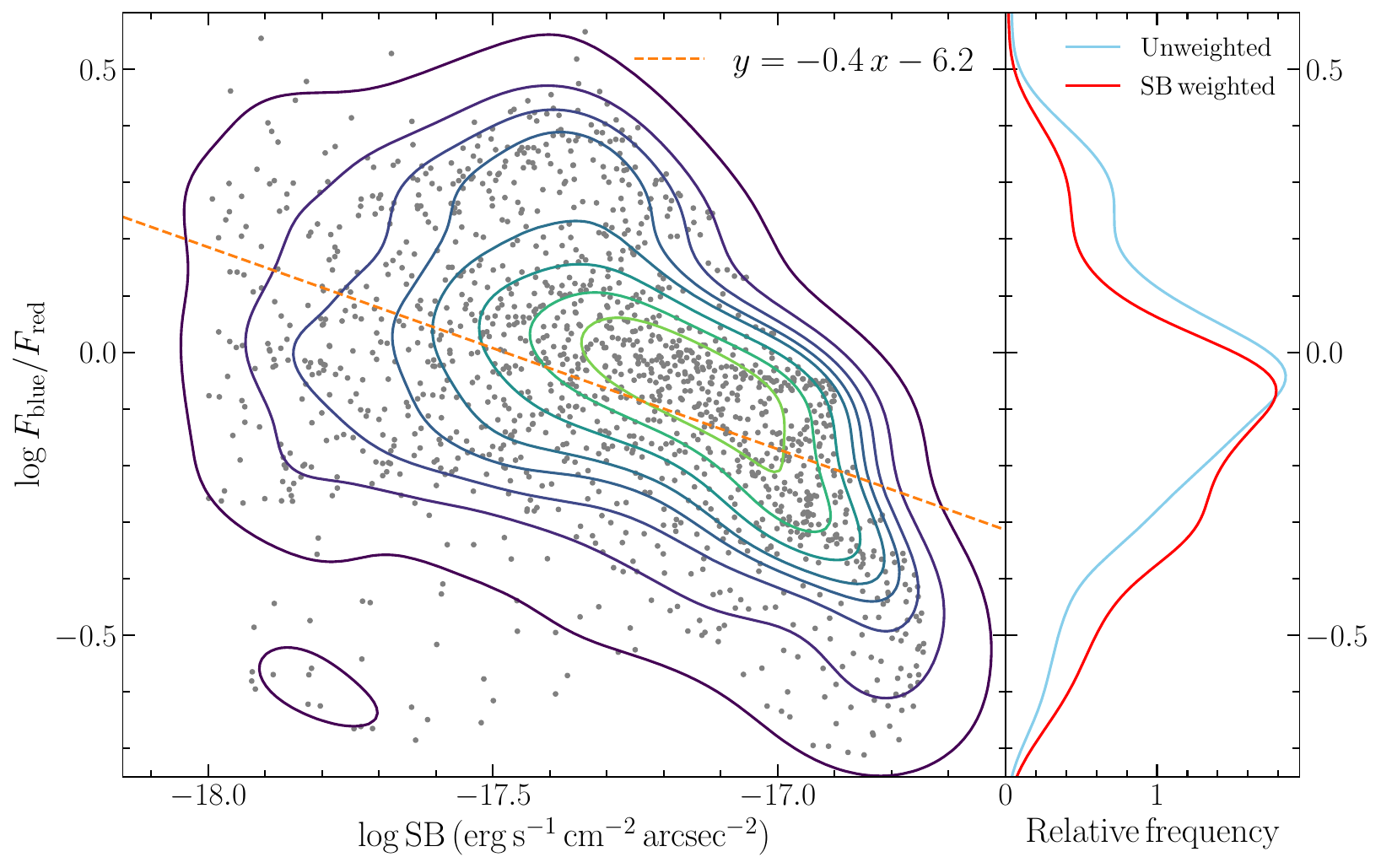}\\
    \caption{The spatial distribution of the blue-to-red flux ratio ($F_{\rm blue}/F_{\rm red}$) and its relation to SB. \emph{Left:} The map of $F_{\rm blue}/F_{\rm red}$ of LAB2 with \lya\ SB contours overlaid (same as in Figure \ref{fig:Lya_images} but in yellow color) and spectrum numbers marked (same as in Figure \ref{fig:Lya_images} but in yellow color for spectra 3 and 14 for clarity). The black contour indicates where $F_{\rm blue}/F_{\rm red}$ = 1. It can be seen that $F_{\rm blue}/F_{\rm red}$ is the lowest in the highest SB region, and increases outwards as the SB decreases. \emph{Right:} The 2D density map of $F_{\rm blue}/F_{\rm red}$ versus SB (on log scale) for all the individual pixels (as shown with grey points) with log\,SB $>$ --18.0 (i.e. within the outermost yellow contour in the left panel). A simple power-law fit yields $F_{\rm blue}/F_{\rm red} \propto \rm SB^{-0.4}$, as shown in the orange dashed line. Note the anti-correlation between $F_{\rm blue}/F_{\rm red}$ and \lya\ SB, which may be due to the decline of the projected line-of-sight outflow velocity and the increase of inflow velocity towards the blob outskirts. Also shown on the right are the unweighted (blue curve) and SB-weighted (red curve) frequency distributions of $F_{\rm blue}/F_{\rm red}$. Both distributions peak at $F_{\rm blue}/F_{\rm red} <$ 1, where the SB-weighted one leans more towards lower $F_{\rm blue}/F_{\rm red}$. 
    \label{fig:ratio_SB}}
\end{figure*}

The diverse \lya\ profiles in three extended \lya\ emitting regions as well as their outskirts are shown in Figure \ref{fig:Lya_linemaps}. All the profiles have been normalised such that the total flux (flux density integrated over the velocities) of the \lya\ line is 1. The small northwest region exhibits a narrow double-peak profile with significant flux between the peaks. Both the northeast and the south region exhibit a similar pattern of an increase in the blue-to-red flux ratio ($F_{\rm blue}/F_{\rm red}$, the flux ratio of the blue peak to the red peak) towards the outer, lower SB regions. We illustrate this quantitatively in Figure \ref{fig:ratio_SB}. 

To calculate the blue-to-red flux ratio of the spectrum of each pixel, we first identified the local minimum (trough) between two peaks of the continuum-subtracted, spatially and spectrally smoothed spectrum, and then integrated both blueward and redward until the flux density goes to zero. The blue-to-red flux ratio is simply the ratio of the integrated fluxes of the blue peak ($F_{\rm blue}$, the \lya\ flux at negative velocities with respect to the local minimum) and the red peak ($F_{\rm red}$, the \lya\ flux at positive velocities with respect to the local minimum). In the left panel of Figure \ref{fig:ratio_SB}, we show the spatial distribution of $F_{\rm blue}/F_{\rm red}$ of LAB2 with SB contours overlaid. It can be seen that $F_{\rm blue}/F_{\rm red}$ is lowest in the highest SB region, and increases outwards as the SB decreases. Regions with $F_{\rm blue}/F_{\rm red}$ > 1 are evident in the outskirts at three different directions: north, southeast and southwest, which is potentially a signature of accreting gas (see e.g. \citealt{Zheng02, Dijkstra06b, Faucher10}).

In the right panel of Figure \ref{fig:ratio_SB}, we show in a 2D density map how $F_{\rm blue}/F_{\rm red}$ varies with SB for all the individual pixels with log\,SB $>$ --18.0 (i.e. within the outermost yellow contour in the left panel). We see that as SB increases, $F_{\rm blue}/F_{\rm red}$ (as shown with grey points) tends to decrease. A simple power-law fit yields $F_{\rm blue}/F_{\rm red} \propto \rm SB^{-0.4}$ (as shown in the orange dashed line). This trend may be due to a combination of: (1) the decline of the projected line-of-sight outflow velocity towards the outskirts of the halo (where the SB is low), assuming a central, roughly symmetric outflow exists (we will quantitatively test this hypothesis in \S\ref{sec:projection}); (2) the increase of inflow velocity at the blob outskirts. Such a transition from outflow to inflow-domination has been observed at $\sim $ 50 kpc for a large sample of star-forming galaxies at $z \sim$ 2 \citep{Chen20}. We also show the unweighted and SB-weighted pixel frequency distributions of $F_{\rm blue}/F_{\rm red}$. Both distributions peak at $F_{\rm blue}/F_{\rm red} <$ 1, where the SB-weighted one leans more towards lower $F_{\rm blue}/F_{\rm red}$. It suggests that our spatially-resolved (not SB-weighted) observations may be better at detecting blue-dominated \lya\ profiles that would be otherwise missed in spatially-integrated (SB-weighted) observations. 

The anti-correlation between the median $F_{\rm blue}/F_{\rm red}$ and SB observed here is similar to the trend observed by \citet{Erb18}, who studied the \lya\ halo of a low-mass star-forming galaxy at $z$ = 2.3 and found that the red peak dominates in the central, high SB region, whereas $F_{\rm blue}/F_{\rm red} \gtrsim 1$ in the outskirts of the halo. They also reported an anti-correlation between $F_{\rm blue}/F_{\rm red}$ and peak separation, which we do not observe in LAB2. Such a difference may reflect the intrinsic difference between \lya\ halos illuminated by a single star-forming galaxy and by potentially multiple sources with various powering mechanisms (as in LAB2). We have also checked several other spectral properties, such as trough position, spectrum width and peak separation, but no significant trends have been found.

\section{Non-detection of Nebular Emission Lines}\label{sec:oiii}
In this section, we summarize the results of our MOSFIRE observations. We have searched through both slits for nebular emission lines and found no significant detection of \oiii\ or \hb\ emission at any location (especially where the \lya\ SB is the highest) on either slit within the region covered by LAB2. For the \oiii$\lambda$5008 line, we measured 2-$\sigma$ flux upper limits of $2.8 \times 10^{-18}$ and $2.5 \times 10^{-18}$ erg\,s$^{-1}$\,cm$^{-2}$ for Slit 1 and 2 respectively, using a window with spatial size of 3\arcsec\ and spectral width of $\sigma$ = 75 km\,s$^{-1}$ assuming a systemic redshift of $z$ = 3.098 (inferred from the following radiative transfer modelling in \S\ref{sec:RT}). We can then utilize these flux upper limits to infer the properties (e.g. star formation rate, SFR) of any possibly existing galaxies along the slits. Taking the observed range of $[\rm O\,{\textsc {iii}}]\lambda$5008/H$\beta \sim 1 - 10$ for $z \sim$ 2 -- 3 star forming galaxies \citep{Steidel14} and assuming zero dust extinction and case B recombination (i.e. H$\alpha$/H$\beta \simeq$ 2.86), we get $[\rm O\,{\textsc {iii}}]\lambda$5008/H$\alpha \sim 0.35 - 3.5$. Using the H$\alpha$ to SFR conversion factor derived by \citet{Kennicutt94} and \citet{Madau98}, the $[\rm O\,{\textsc {iii}}]\lambda$5008 2-$\sigma$ upper limits correspond to SFR $\sim$ 0.6 -- 6 M$_{\odot}$ / yr.

Although the non-detection of the nebular emission lines is puzzling, it is clear that the \lya\ profiles near the high SB center are red-dominated and suggest the presence of outflows, which is most likely due to star formation or AGN-driven winds. It is therefore reasonable to hypothesize the existence of star-forming galaxies or AGNs hidden by dust extinction and/or contamination of a foreground source (see the right panel of Figure \ref{fig:Lya_images} for the location of a possible low-$z$ interloper) near the \lya\ SB peak. Our subsequent radiative transfer modelling analysis is also based on this assumed `central powering + scattering' scenario\footnote{One may speculate that the `cold accretion' scenario can also produce red-dominated profiles if aided by IGM absorption preferentially on the blue side. However, such a scenario is likely to result in multiple (e.g. triple) peaks at $z \sim$ 3, which we have not observed in LAB2 \citep{Byrohl2020}}.

\section{Radiative Transfer modelling Using The Multiphase Clumpy Model}\label{sec:RT}

\subsection{Methodology}
To decode physical properties of the gas in LAB2 from the observational data, we used radiative transfer modelling to fit the spatially-resolved \lya\ profiles. Similar to \citet{Li21}, we adopted a multiphase, clumpy model, which assumes cool, spherical \HI\ clumps moving within a hot, ionized inter-clump medium (ICM) \citep{Laursen13,Gronke16_model}. The most crucial parameters of this model include: (1) the cloud covering factor ($f_{\rm {\rm cl}}$), which is the mean number of clumps per line-of-sight from the center to the boundary of the simulation sphere; (2) the residual\footnote{Here we have assumed that the hydrogen in the hot ICM is highly ionized and only a very small fraction exists in the form of \HI\ (i.e. $x_{\rm HI,\,{\rm ICM}} \ll 1$).} \HI\ number density in the ICM ($n_{\rm HI,\,{\rm ICM}}$), which determines the depth of the absorption trough; (3) the kinematics of the clumps, consisting of an isotropic Gaussian random motion (characterized by $\sigma_{\rm {\rm cl}}$, the velocity dispersion of the clumps) and a symmetric radial outflow with a constant velocity $v_{\rm cl}$; (4) the $\rm H\,{\textsc {i}}$ column density of the clumps.

Note that \lya\ radiative transfer in such a multiphase, clumpy medium exhibits two characteristic regimes defined by the values of $f_{\rm cl}$ \citep{Gronke16, Gronke17b}. If $f_{\rm cl}$ is much larger than a critical value $f_{\rm cl, crit}$ (which is a function of the \HI\ column density and kinematics of the clumps, see Appendix B of \citet{Li21} for a detailed derivation), the photons would escape as if the medium is homogeneous and the emergent spectra are similar to those predicted by the `shell  model' \citep{Gronke17b}. Otherwise, for a moderate number of clumps per line-of-sight, the photons preferentially travel in the hot ionized ICM and escape close to the \lya\ line center. 

\begin{table}
    \centering
    \scriptsize \caption{Parameter values of the grids of models.}
    \label{tab:params}
    \setlength{\tabcolsep}{3pt}
    \begin{tabular}{cccc}
    \hline\hline
    Model & Parameter & Definition & Values\\ 
     (1)  & (2) & (3) & (4) \\
    \hline
    & ${\rm log}\,n_{\rm HI,\,{\rm ICM}}$ & ICM \HI\ number density & (-7.5, -6.5, -5.5, -4.5) log cm$^{-3}$\\
     & $F_{\rm V}$ & Volume filling factor & (0.02, 0.1, 0.2, 0.3, 0.4, 0.5, 0.6)\\
     & ${\rm log}\,N_{\rm HI,\,cl}$ & Clump {\rm H\,{\textsc {i}}} column density & (17, 17.5, 18, 18.5, 19) log cm$^{-2}$\\
    Clumpy & $\sigma_{\rm cl}$ & Clump velocity dispersion & (0, 100, \ldots, 600) km\,s$^{-1}$ \\
     & $v_{\rm cl}$ & Clump outflow velocity & (0, 100, 200, 300) km\,s$^{-1}$ \\
     & $\Delta v_{\rm clumpy}$ & Velocity shift & [-200,\,200] km\,s$^{-1}$ \\
     \hline
     & $v_{\rm exp}$ & Shell expansion velocity & (0, 2, 5, 8, 10, 15, 20, 30, \ldots, 490) km\,s$^{-1}$\\
     & $\log N_{\rm HI, shell}$ & Shell \HI\ column density & (16.0, 16.2, \ldots, 21.8) log cm$^{-2}$\\
    Shell & $\log T_{\rm shell}$ & Shell (effective) temperature & (4.0, 4.4, \ldots, 5.8) log K \\
     & $\sigma_{\rm i}$ & Intrinsic spectrum width & [1,\,800] km\,s$^{-1}$ (continuous) \\
     & $\Delta v_{\rm shell}$ & Velocity shift & [-200,\,200] km\,s$^{-1}$ (continuous)\\
    \hline\hline
    \end{tabular}
    \begin{tablenotes}
    \item \textbf{Notes.} The parameter values of the model grids that we used for fitting the \lya\ profiles. The columns are: (1) model type; (2) parameter name; (3) definition of the parameter; (4) parameter values on the grid. Note that negative values for $v_{\rm cl}$ and $v_{\rm exp}$ are also allowed in the fitting.
    \end{tablenotes}
\end{table}

\begin{figure*}
\centering
\includegraphics[width=0.322\textwidth]{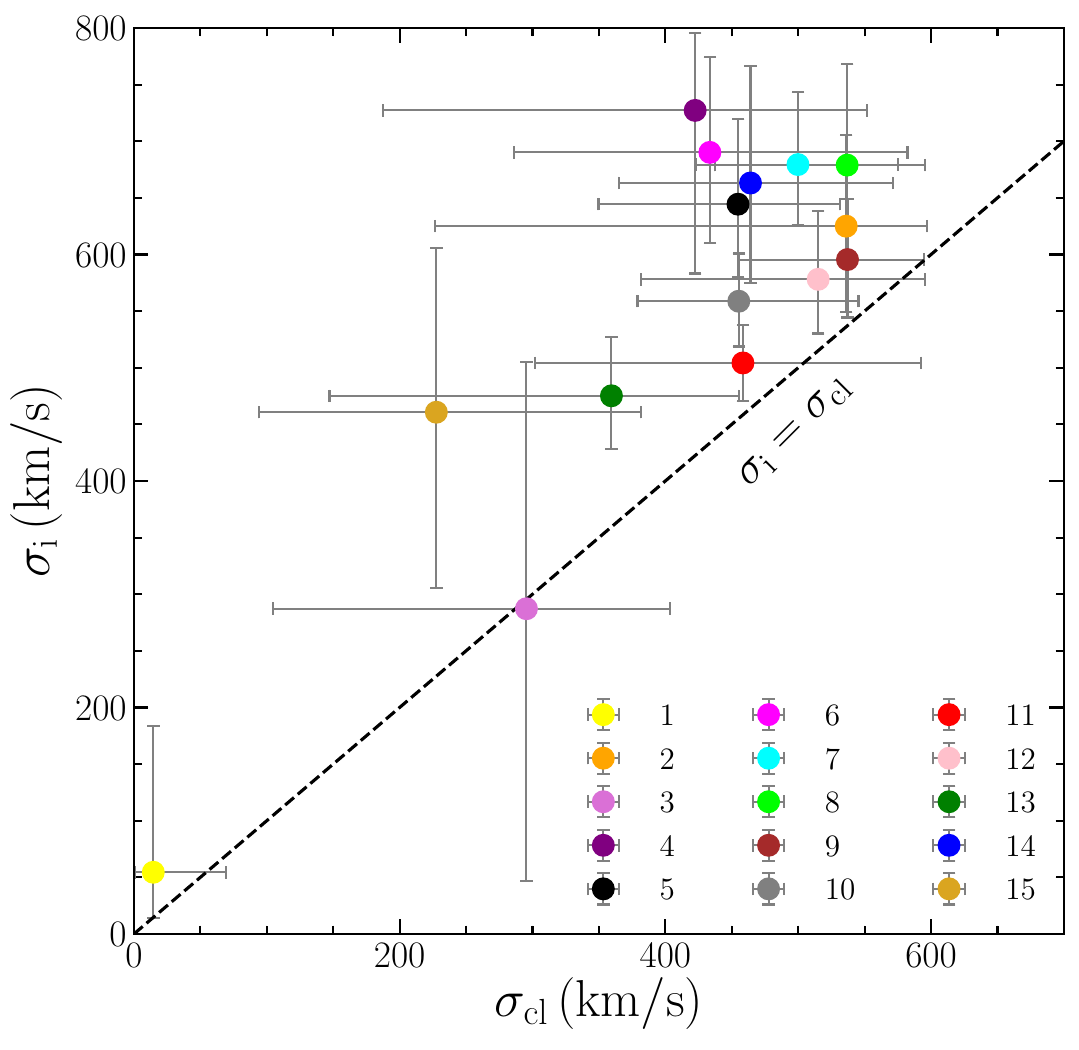}
\includegraphics[width=0.33\textwidth]{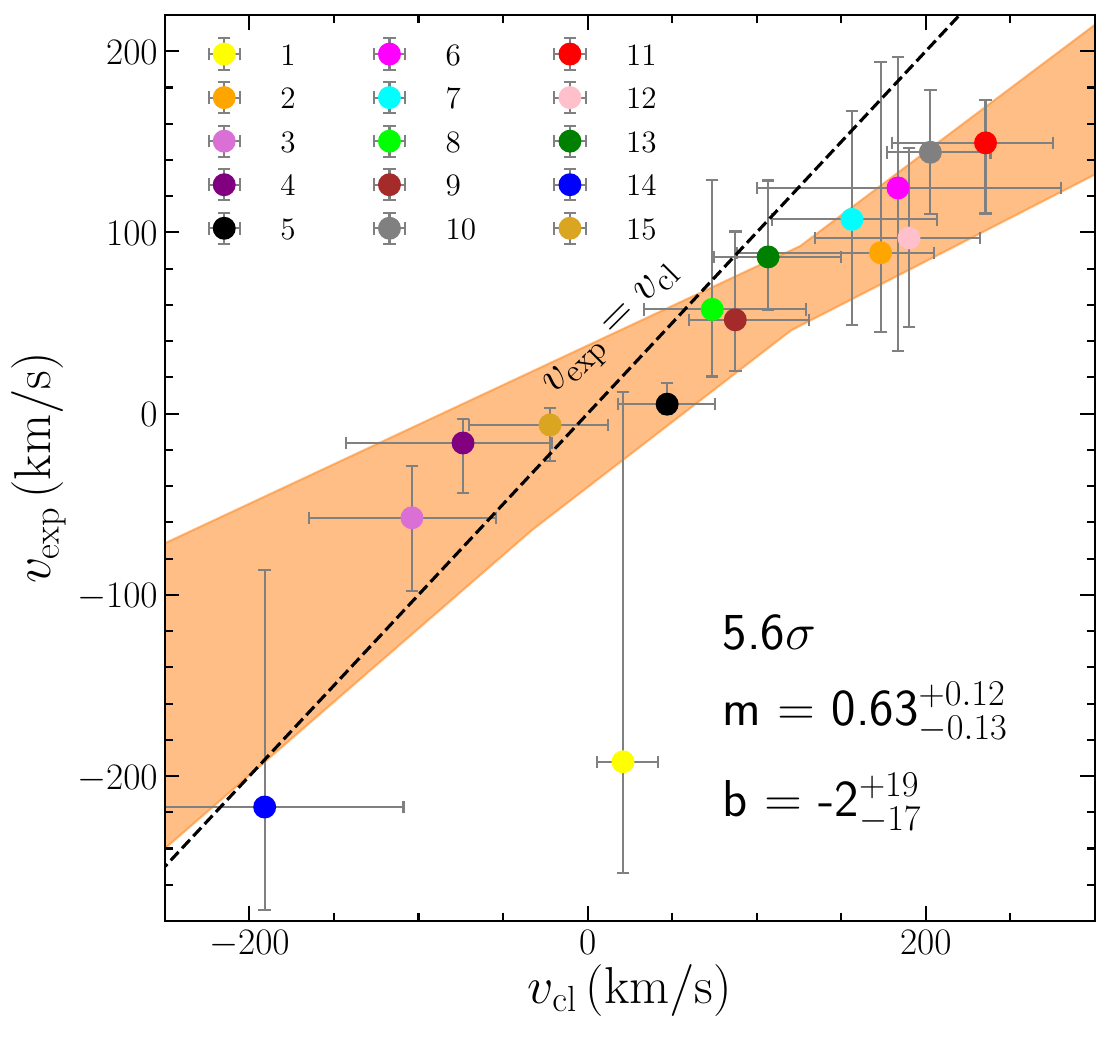}
\includegraphics[width=0.33\textwidth]{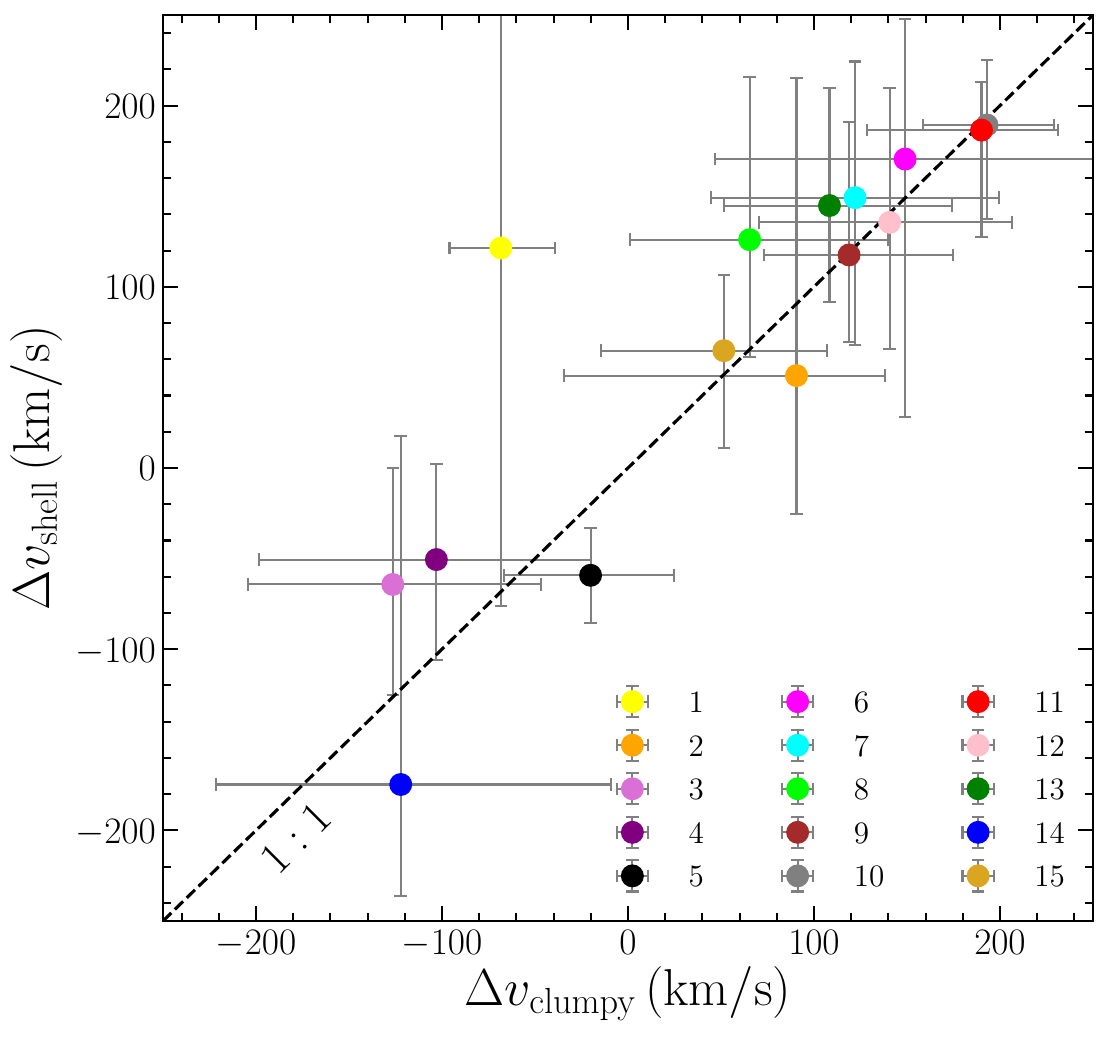}
    \caption{Correlations between the multiphase, clumpy model parameters and the shell model parameters. Only one significant ($\gtrsim5\sigma$) positive correlation is observed between $v_{\rm cl}$ and $v_{\rm exp}$. Interestingly, $\sigma_{\rm cl} \leq \sigma_{\rm i}$ and $|v_{\rm cl}| \geq |v_{\rm exp}|$ are almost always true. The $\Delta v_{\rm clumpy}$ v.s. $\Delta v_{\rm shell}$ correlation is insignificant, but the data points are broadly consistent with a 1-to-1 relation within 2$\sigma$ uncertainties. The color-coded points represent different \lya\ spectra (see Figure \ref{fig:Lya_images} and \ref{fig:Lya_linemaps}). The 2-$\sigma$ uncertainties of the data points are indicated by grey error bars. For the $v_{\rm cl}$ v.s. $v_{\rm exp}$ correlation, the level of significance and the linear best-fit coefficients (slope $m$ and intercept $b$, with 1-$\sigma$ uncertainties) are shown at the lower right corner in the middle panel. The orange shaded region represent the range of twenty best-fits of the data points perturbed by their uncertainties.
    \label{fig:correlation}}
\end{figure*}

\begin{table*}
\renewcommand\arraystretch{1.8}
  \scriptsize \caption{Fitted parameters and derived quantities of the multiphase, clumpy model and the shell model.}
  \label{tab:Fitting_results}
  \centering
  \begin{adjustbox}{angle=0}
  \setlength{\tabcolsep}{3pt}
  \begin{tabular}{ccc|cccccc|cc|ccccc}%
\hline%
\hline%
\multicolumn{3}{c|}{}&\multicolumn{6}{c|}{Clumpy Model Parameters}&\multicolumn{2}{c|}{Derived Quantities}&\multicolumn{5}{c}{Shell Model Parameters}\\%
\hline%
No.&RA (J2000)&Dec (J2000)&${\rm log}\,n_{\rm HI,\,{\rm ICM}}$&$F_{\rm V}$&${\rm log}\,N_{\rm HI,\,{\rm cl}}$&$\sigma_{\rm cl}$&$v_{\rm cl}$&$\Delta v_{\rm clumpy}$&$f_{\rm cl}/f_{\rm cl, crit}$&$\log\tau_{\rm 0, ICM}$&$v_{\rm exp}$&${\rm log}\,N_{\rm HI,\,{\rm shell}}$&${\rm log}\,T_{\rm shell}$&$\sigma_{\rm i}$&$\Delta v_{\rm shell}$\\%
&&&(cm$^{-3})$&&(cm$^{-2})$&(km\,s$^{-1})$&(km\,s$^{-1})$&(km\,s$^{-1})$&&&(km\,s$^{-1})$&(cm$^{-2})$&(K)&(km\,s$^{-1})$&(km\,s$^{-1})$\\%
(1)&(2)&(3)&(4)&(5)&(6)&(7)&(8)&(9)&(10)&(11)&(12)&(13)&(14)&(15)&(16)\\%
\hline%
1&22:17:39.179&+00:13:33.17&-7.3$^{+0.3}_{-0.2}$&0.31$^{+0.26}_{-0.23}$&17.4$^{+0.6}_{-0.4}$&14$^{+54}_{-13}$&20$^{+20}_{-15}$&-68$^{+20}_{-23}$&23.3$^{+20.0}_{-18.4}$&0.7$^{+0.3}_{-0.2}$&-192$^{+204}_{-61}$&19.0$^{+0.4}_{-2.7}$&4.9$^{+0.8}_{-0.8}$&54$^{+129}_{-40}$&121$^{+44}_{-188}$\\%
2&22:17:38.919&+00:13:33.47&-4.7$^{+0.2}_{-0.6}$&0.07$^{+0.37}_{-0.04}$&18.1$^{+0.5}_{-0.9}$&536$^{+61}_{-309}$&173$^{+31}_{-85}$&90$^{+35}_{-91}$&0.6$^{+8.2}_{-0.3}$&3.3$^{+0.2}_{-0.6}$&88$^{+105}_{-43}$&19.7$^{+0.5}_{-1.6}$&5.6$^{+0.1}_{-1.4}$&624$^{+80}_{-75}$&50$^{+125}_{-63}$\\%
3&22:17:38.859&+00:13:31.97&-4.7$^{+0.2}_{-0.6}$&0.34$^{+0.23}_{-0.24}$&17.8$^{+1.1}_{-0.7}$&295$^{+107}_{-190}$&-103$^{+49}_{-60}$&-126$^{+62}_{-49}$&5.2$^{+11.3}_{-3.6}$&3.3$^{+0.2}_{-0.6}$&-57$^{+28}_{-40}$&20.3$^{+0.4}_{-0.4}$&5.6$^{+0.2}_{-0.6}$&287$^{+217}_{-240}$&-64$^{+57}_{-45}$\\%
4&22:17:38.999&+00:13:31.37&-5.4$^{+0.8}_{-1.4}$&0.32$^{+0.25}_{-0.21}$&18.0$^{+0.9}_{-0.9}$&422$^{+129}_{-234}$&-73$^{+52}_{-69}$&-102$^{+64}_{-65}$&3.8$^{+8.1}_{-2.5}$&2.6$^{+0.8}_{-1.4}$&-16$^{+13}_{-27}$&20.8$^{+0.5}_{-0.4}$&5.0$^{+0.6}_{-0.9}$&727$^{+68}_{-144}$&-50$^{+51}_{-48}$\\%
5&22:17:39.059&+00:13:29.27&-5.7$^{+0.4}_{-0.4}$&0.26$^{+0.17}_{-0.16}$&17.4$^{+1.1}_{-0.4}$&454$^{+76}_{-105}$&47$^{+28}_{-29}$&-20$^{+34}_{-36}$&2.6$^{+2.0}_{-1.4}$&2.3$^{+0.4}_{-0.4}$&5$^{+11}_{-5}$&20.6$^{+0.4}_{-0.4}$&4.8$^{+0.6}_{-0.7}$&644$^{+74}_{-64}$&-59$^{+23}_{-25}$\\%
6&22:17:38.879&+00:13:28.37&-5.9$^{+1.1}_{-0.7}$&0.10$^{+0.33}_{-0.08}$&18.5$^{+0.5}_{-1.4}$&433$^{+148}_{-147}$&183$^{+96}_{-83}$&148$^{+45}_{-59}$&1.3$^{+3.7}_{-1.1}$&2.0$^{+1.1}_{-0.7}$&124$^{+72}_{-90}$&19.5$^{+0.8}_{-0.3}$&4.6$^{+0.5}_{-0.6}$&690$^{+84}_{-80}$&170$^{+28}_{-109}$\\%
7&22:17:38.959&+00:13:28.07&-5.9$^{+0.6}_{-0.4}$&0.23$^{+0.17}_{-0.17}$&17.6$^{+1.1}_{-0.5}$&499$^{+75}_{-76}$&156$^{+50}_{-47}$&122$^{+58}_{-60}$&2.1$^{+1.5}_{-1.5}$&2.1$^{+0.6}_{-0.4}$&107$^{+59}_{-58}$&19.4$^{+0.4}_{-0.3}$&4.6$^{+0.4}_{-0.5}$&679$^{+63}_{-53}$&149$^{+45}_{-56}$\\%
8&22:17:39.179&+00:13:27.77&-6.4$^{+0.4}_{-0.3}$&0.18$^{+0.13}_{-0.12}$&17.5$^{+1.1}_{-0.4}$&536$^{+58}_{-99}$&73$^{+55}_{-40}$&65$^{+50}_{-49}$&1.6$^{+1.1}_{-0.9}$&1.5$^{+0.4}_{-0.3}$&57$^{+71}_{-37}$&17.7$^{+1.4}_{-1.5}$&4.9$^{+0.5}_{-0.7}$&678$^{+89}_{-75}$&126$^{+54}_{-52}$\\%
9&22:17:39.019&+00:13:27.17&-6.2$^{+0.3}_{-0.3}$&0.14$^{+0.08}_{-0.06}$&17.5$^{+0.7}_{-0.4}$&537$^{+57}_{-81}$&87$^{+43}_{-27}$&118$^{+35}_{-36}$&1.2$^{+0.7}_{-0.5}$&1.7$^{+0.3}_{-0.3}$&51$^{+48}_{-28}$&19.4$^{+0.5}_{-0.6}$&4.7$^{+0.3}_{-0.7}$&595$^{+53}_{-51}$&117$^{+54}_{-39}$\\%
10&22:17:38.959&+00:13:26.27&-6.2$^{+0.3}_{-0.3}$&0.14$^{+0.12}_{-0.06}$&17.8$^{+0.4}_{-0.7}$&455$^{+90}_{-76}$&202$^{+35}_{-25}$&193$^{+6}_{-22}$&1.4$^{+1.0}_{-0.7}$&1.8$^{+0.3}_{-0.3}$&144$^{+34}_{-34}$&19.3$^{+0.2}_{-0.1}$&4.7$^{+0.3}_{-0.6}$&558$^{+42}_{-39}$&189$^{+10}_{-38}$\\%
11&22:17:38.899&+00:13:25.37&-5.7$^{+0.5}_{-0.8}$&0.07$^{+0.16}_{-0.03}$&17.9$^{+0.3}_{-0.5}$&458$^{+134}_{-156}$&235$^{+40}_{-55}$&190$^{+9}_{-27}$&0.6$^{+1.9}_{-0.4}$&2.3$^{+0.5}_{-0.8}$&149$^{+23}_{-38}$&19.4$^{+0.3}_{-0.1}$&4.8$^{+0.2}_{-0.7}$&504$^{+33}_{-33}$&186$^{+12}_{-44}$\\%
12&22:17:38.999&+00:13:24.77&-5.5$^{+0.7}_{-0.6}$&0.07$^{+0.17}_{-0.04}$&18.1$^{+0.4}_{-1.0}$&515$^{+80}_{-133}$&190$^{+42}_{-55}$&140$^{+50}_{-43}$&0.7$^{+1.7}_{-0.4}$&2.4$^{+0.7}_{-0.6}$&96$^{+49}_{-49}$&19.6$^{+0.4}_{-0.4}$&4.9$^{+0.4}_{-0.8}$&578$^{+60}_{-47}$&135$^{+54}_{-49}$\\%
13&22:17:39.059&+00:13:22.97&-5.8$^{+0.4}_{-0.4}$&0.14$^{+0.20}_{-0.07}$&17.4$^{+0.9}_{-0.4}$&359$^{+95}_{-212}$&106$^{+43}_{-31}$&108$^{+49}_{-47}$&1.8$^{+7.4}_{-0.9}$&2.2$^{+0.4}_{-0.4}$&86$^{+42}_{-29}$&19.3$^{+0.4}_{-0.8}$&5.2$^{+0.3}_{-0.4}$&475$^{+52}_{-46}$&144$^{+49}_{-44}$\\%
14&22:17:39.239&+00:13:22.37&-6.2$^{+0.9}_{-0.5}$&0.23$^{+0.21}_{-0.19}$&17.6$^{+1.2}_{-0.6}$&464$^{+107}_{-99}$&-190$^{+82}_{-77}$&-122$^{+77}_{-62}$&2.2$^{+2.0}_{-1.8}$&1.8$^{+0.9}_{-0.5}$&-217$^{+130}_{-56}$&18.5$^{+1.2}_{-2.3}$&5.1$^{+0.4}_{-0.8}$&663$^{+103}_{-88}$&-174$^{+140}_{-24}$\\%
15&22:17:38.879&+00:13:22.67&-6.5$^{+0.8}_{-0.9}$&0.28$^{+0.28}_{-0.23}$&17.9$^{+1.0}_{-0.8}$&227$^{+153}_{-133}$&-22$^{+34}_{-47}$&51$^{+43}_{-45}$&6.3$^{+13.5}_{-5.4}$&1.4$^{+0.8}_{-0.9}$&-6$^{+9}_{-19}$&20.3$^{+0.4}_{-1.1}$&4.7$^{+0.7}_{-0.7}$&460$^{+145}_{-155}$&64$^{+40}_{-49}$\\%
\hline%
\hline%
\end{tabular}
  \end{adjustbox}
  \begin{tablenotes}
  \small
    \item \textbf{Notes.} Fitted parameters (averages and 2.5\% -- 97.5\% quantiles, i.e. 2-$\sigma$ confidence intervals) of the multiphase clumpy model and shell model, and derived quantities. The columns are: (1) the spectrum number (as marked in Figure \ref{fig:Lya_linemaps}); (2) the right ascension of the center of the extracted region; (3) the declination of the center of the extracted region; (4) the \HI\ number density in the ICM; (5) the cloud volume filling factor; (6) the clump $\rm H\,{\textsc {i}}$ column density; (7) the velocity dispersion of the clumps; (8) the radial outflow velocity of the clumps; (9) the velocity shift relative to the initial guess for the systemic redshift ($z$ = 3.09 for spectrum 1 and $z$ = 3.098 for all other spectra; a negative/positive value means that the best-fit model spectrum has been blue/redshifted to match the data); (10) the clump covering fraction (defined as the number of clumps per line-of-sight) normalised by the critical clump covering fraction. In our case $f_{\rm cl}$ = 75\,$F_{\rm V}$. The critical clump covering fraction, $f_{\rm cl, crit}$, determines different physical regimes and is calculated via Eq. (B3) in \citet{Li21}; (11) the optical depth at the \lya\ line center of the \HI\ in the ICM; (12) the expansion velocity of the shell; (13) the \HI\ column density of the shell; (14) the \HI\ temperature of the shell; (15) the required intrinsic line width; (16) the velocity shift relative to the initial guess for the systemic redshift.
   \end{tablenotes}
\end{table*}

We constructed a five-dimensional hypercubic grid of models by varying the aforementioned five crucial physical parameters: [${\rm log}\,n_{\rm HI,\,{\rm ICM}}, F_{\rm V}, {\rm log}\,N_{\rm HI,\,{\rm cl}}, \sigma_{\rm cl}, v_{\rm cl}$]\footnote{For convenience we varied $F_{\rm V}$ rather than $f_{\rm cl}$ when generating clumps. These two parameters are proportionally related via the relation $f_{\rm cl}$ = 3$r_{\rm gal}$/4$r_{\rm cl}$\,$F_{\rm V}$, where $r_{\rm gal}$ = 5\,kpc is the radius of the simulation sphere and $r_{\rm cl}$ = 50\,pc is the clump radius (hence $f_{\rm cl}$ = 75\,$F_{\rm V}$ in our case). Note that the physical sizes used in the simulation are unimportant to \lya\ spectra; it is the ${\rm H\,{\textsc {i}}}$ column densities and corresponding optical depths that are important and can actually be constrained by the data.}. The prior ranges of the parameters are summarized in Table \ref{tab:params}. We fixed the subdominant parameters, such as the clump temperature $T_{\rm cl}$ to 10$^{4}$\,K and the ICM temperature $T_{\rm ICM}$ to 10$^{6}$\,K\footnote{Different from \citet{Li21}, we set the ICM co-outflow with the clumps (i.e. $v_{\rm cl} = v_{\rm ICM}$), as both $v_{\rm cl}$ and $v_{\rm ICM}$ decrease the blue-to-red flux ratio, and in reality, a large velocity difference between two different phases would likely destroy the clumps via hydrodynamical instabilities on short timescales.}. We varied an additional parameter, $\Delta v_{\rm clumpy}$, continuously in post-processing. This $\Delta v_{\rm clumpy}$ parameter represents the best-fit systemic redshift of the \lya\ source function relative to $z$ = 3.098 (the initial guess for the systemic redshift -- it is where the trough of most \lya\ spectra is located)\footnote{Note that the northeast \lya\ emitting region is located at a lower redshift, $z$ = 3.09, which we adopted as the initial guess for spectrum 1.}. Furthermore, we considered \textit{inflow} velocities (i.e. $v_{\rm cl} < 0$) by mirroring the model spectra with respect to the line center. 

Such a configuration amounts to 3920 models in total. Each model is calculated via radiative transfer using $10000$ \lya\ photon packages generated from a Gaussian intrinsic spectrum \emph{N}(0, $\sigma^2_{\rm i, cl}$), where $\sigma_{\rm i, cl}$ = 12.85\,km\,s$^{-1}$ is the canonical thermal velocity dispersion of $T$ = 10$^4$\,K gas. To properly explore the multimodal posterior of the parameters, we used a python nested sampling package \texttt{dynesty} \citep{Skilling04, Skilling06, Speagle20} for our fitting pipeline.

\subsection{Results}\label{sec:results}
\subsubsection{Fits \& Derived Parameters}\label{sec:correlations}

We selected fifteen representative \lya\ spectra from the high SB regions in LAB2 for further model fitting. The positions of these spectra are shown in Figure \ref{fig:Lya_images}, and the profiles are shown in Figure \ref{fig:Lya_linemaps}. All spectra were extracted from certain single spatial pixels of the boxcar-smoothed KCWI datacube, each representing the average spectrum of a corresponding 3 pixel x 3 pixel boxcar-smoothed region.

In addition to using the multiphase, clumpy models to fit the \lya\ spectra, we also adopted the widely used `shell model' (e.g. \citealt{Ahn03,Verhamme06,Dijkstra06b}). A similar three-dimensional cubic grid of shell models was constructed by varying three parameters: the shell expansion velocity $v_{\rm exp}$, the shell \HI\ column density ${\rm log}\,N_{\rm HI,\,{\rm shell}}$, and the effective temperature of \HI\ in the shell, ${\rm log}\,T_{\rm shell}$. Two more parameters, namely the intrinsic line width $\sigma_{\rm i}$ and the velocity shift with respect to the initial guess for the systemic redshift ($z$ = 3.09 for spectrum 1 and $z$ = 3.098 for all other spectra), $\Delta v_{\rm shell}$, are varied continuously in post-processing\footnote{The detailed configuration of the shell model is presented in \citet{Gronke15}, and an example of fitting \lya\ spectra with a grid of shell models is presented, for instance, in \citet{Gronke17}.}. 

During the fitting procedure, each model spectrum is calculated via linear flux interpolation on the model grid\footnote{Interpolation is necessary here because it is too computationally expensive if in the nested sampling process we calculate the model spectrum at each point in the parameter space “on the fly” (i.e. performing RT at that point). So instead, we used a pre-calculated and saved grid of models to calculate each model spectrum via a linear flux interpolation of the adjacent grid models, which is accurate enough for our purpose.} and is convolved with the KCWI line-spread function (LSF, a Gaussian with $\sigma$ = 65 km\,s$^{-1}$) before being compared to the observed \lya\ profiles. To better reproduce the profiles dominated by a blue peak (e.g. spectra 3, 4, 14 and 15), we have also incorporated model spectra with negative $v_{\rm cl}$ or $v_{\rm exp}$ that have been `mirrored' in the velocity space from their positive $v_{\rm cl}/v_{\rm exp}$ counterparts into our calculation. The best-fit model spectra are also shown in Figure \ref{fig:Lya_linemaps}.

In Figure \ref{fig:Lya_linemaps}, one can see that, in most cases, both the multiphase, clumpy model fits and shell model fits match the observed \lya\ profiles reasonably well. The values of the fitted parameters are presented in Table \ref{tab:Fitting_results}, and the derived joint and marginal posterior probability distributions are presented in Appendix \ref{sec:posterior}. The best-fit parameters are determined as the highest likelihood point in the sampled parameter space, and the uncertainties in the fitted parameters are determined as certain quantiles (e.g. 2.5\% -- 97.5\%, or 2-$\sigma$ confidence intervals) of the samples in the marginalized posterior probability distributions. Note that the best-fit parameters of the multiphase, clumpy model derived here should be interpreted as the the local gas properties -- e.g. ${\rm H\,{\textsc {i}}}$ column densities, clump velocity dispersions, and in particular, clump outflow velocities {\it{along the line-of-sight}} relative to the local systemic redshift. Such a practice provides the distribution of gas kinematics and {\rm H\,{\textsc {i}}} column densities without relying on the assumption of the location of the \lya\ source (see Appendix \ref{sec:justification}).

Motivated by the fact that certain parameters from both models should control similar \lya\ spectral properties, we further attempted to find the link between the two models by correlating parameter pairs. As a result, we have only observed one significant ($\gtrsim5\sigma$) positive correlation between $v_{\rm cl}$ and $v_{\rm exp}$. We have also performed linear regressions to this correlation and estimated the uncertainties in the coefficients (slope $m$ and intercept $b$) by perturbing the data points with asymmetric Gaussian noise with amplitude proportional to the error bars. The results are shown in Figure \ref{fig:correlation}.

The existence of the significant correlation between $v_{\rm cl}$ and $v_{\rm exp}$ can be easily understood since this parameter pair controls the same spectral property, namely the blue-to-red flux ratio. Interestingly, we see that $\sigma_{\rm cl} \leq \sigma_{\rm i}$ and $|v_{\rm cl}| \geq |v_{\rm exp}|$ are almost always true. These results naturally alleviated the tension between the fitted shell model parameters and the observational constraints reported in e.g. \citet{Orlitova18}, namely: (1) the required $\sigma_{\rm i}$ are on average three times broader than the observed non-resonant Balmer lines; (2) the derived $v_{\rm exp}$ are smaller than the outflow velocities determined from UV absorption lines. This suggests that the photon scattering between randomly moving clumps may be an efficient way of broadening \lya\ profiles and circumventing overlarge $\sigma_{\rm i}$ (see also  \citealt{Hashimoto15}), and that $v_{\rm cl}$ in the clumpy medium is less efficient at increasing the spectrum asymmetry than $v_{\rm exp}$. Such distinctions reflect the intrinsic differences between two models: in the shell model, all the photons have to traverse the shell and thus are shaped by the same shell outflow/inflow velocity; whereas in the multiphase, clumpy model, the photons can randomly walk between the clumps or even diffuse outwards with their frequencies unaffected\footnote{Note that we focused on the $f_{\rm cl}/f_{\rm cl,crit} \simeq 1 - 10$ regime in this work.} \citep{Neufeld91,Gronke16_model}. This means that the effective outflow velocity `experienced' by the photons in the multiphase, clumpy model is smaller than that in the shell model, which needs to be compensated by a larger $v_{\rm cl}$. Furthermore, as scattering orthogonally off the flowing clumps may yield additional broadening to the spectra \citep{Li21}, the $\sigma_{\rm cl}$ values required to achieve the large observed widths of \lya\ profiles are lower.

By the same token, correlations between (1) $n_{\rm HI,\,{\rm ICM}}$ (and $F_{\rm V}N_{\rm HI,\,{\rm cl}}$\footnote{Both $n_{\rm HI,\,{\rm ICM}}$ and $F_{\rm V}N_{\rm HI,\,{\rm cl}}$ (or $f_{\rm cl}N_{\rm HI,\,{\rm cl}}$) contribute to the effective column density in the multiphase, clumpy model (see Eq. (1) in \citealt{Gronke16_model}).}) and $N_{\rm HI,\,{\rm shell}}$; (2) $\Delta v_{\rm clumpy}$ and $\Delta v_{\rm shell}$ may exist, as they control the peak separation and the systemic redshift of the \lya\ source, respectively. However, we found that the former correlation is insignificant, as the best-fits from two models prefer different peak separations in many cases (see e.g. spectra 7, 8, 9 and 13). The latter correlation is also insignificant ($\lesssim 2\sigma$) with one apparent outlier -- for spectrum 1, the multiphase, clumpy model prefers outflow, whereas the shell model prefers inflow. However, all 15 data points are broadly consistent with a 1-to-1 relation within 2$\sigma$ uncertainties. Such a correspondence between $\Delta v_{\rm clumpy}$ and $\Delta v_{\rm shell}$ suggests that the inconsistency reported in \citet{Orlitova18}, i.e.  the best-fit systemic redshifts inferred from the shell model are larger by 10 – 250 km\,s$^{-1}$ than those determined from optical emission lines, also exists in the multiphase, clumpy model. Further observations of non-resonant lines that are available for other objects are necessary to solve this issue.

We have not found any straightforward analytic mapping functions that can directly convert the best-fit parameters from one model to the other. In our future work, we will explore whether such analytic mapping functions exist between certain parameters first in the $f_{\rm cl} \gg f_{\rm cl, crit}$ regime, and then in the regime that we have explored in this work, i.e. $f_{\rm cl}$ is higher than $f_{\rm cl, crit}$ but mostly within one order of magnitude, where only qualitative trends between parameter pairs have been observed.

Although the shell model fits have comparable likelihoods to the multiphase, clumpy model fits, they are less likely to be informative of the actual physical conditions in the circumgalactic medium (CGM), because (1) the shell model only has a single phase of \HI\ with low-to-medium effective temperature ($T_{\rm shell} \sim 10^{4-5} \rm K$); (2) the aforementioned $v_{\rm exp}, \sigma_{\rm i}$ and $\Delta v_{\rm shell}$ values that are inconsistent with the observational constraints (e.g. \citealt{Orlitova18}). Whereas the multiphase, clumpy model is not only likely a more realistic description of the CGM \citep{Tumlinson17}, but it also yields more reasonable physical parameters (as we will elaborate below). Therefore, we will focus on the multiphase, clumpy model in the rest of this paper.

\subsubsection{Interpretation of Fitted Parameters}
In this section, we discuss how realistic the fitted parameters of the multiphase, clumpy model are compared to other studies.

\begin{enumerate}

\item \emph{Covering fraction of the cool clumps}: The derived volume filling factors ($F_{\rm V}$) range from $\sim$ 0.1 -- 0.5, which convert to covering factors ($f_{\rm cl}$) of $\sim$ 7 -- 40 for clumps with $N_{\rm HI,\,{\rm cl}} \gtrsim 10^{17} \rm cm^{-2}$. Such high covering factors effectively correspond to covering \emph{fractions} of unity\footnote{This is true especially in the central region; in the outskirts the covering fraction may decrease even for a homogeneous clump distribution (as assumed in this work). Technically, the effective covering factor to an external observer is $\tilde{f_{\rm cl}} = f_{\rm cl} \frac{2\sqrt{R^2 - b^2}}{R}$, which corresponds to an (area) covering fraction of $f_{\rm A}(b) = 1 - \exp(\tilde{f_{\rm cl}})$, i.e. one minus the Possion probability of photons intersecting with zero clumps. Here $R$ is the radius of the studied region, and $b$ is the projected distance from the center of the region relative to the line-of-sight.} \citep{Laursen13}. This result is consistent with the recent findings in \citet{Wisotzki18}, where they observed low SB \lya\ emission surrounding high-$z$ faint galaxies with MUSE and claimed that the \HI\ covering fractions around galaxies should be sufficiently close to unity at $z > 3$, assuming the spatial distribution of circumgalactic \HI\ is similar to the \lya-emitting gas. Additionally, high \HI\ covering fractions around galaxies have also been observed at lower redshifts ($z < 3$, see e.g. \citealt{Chen01, Adelberger03, X11, Rudie12, Tumlinson13}), suggesting that large \HI\ covering fractions should be universally present across different cosmic epochs. 

\item \emph{Velocity of the cool clumps}: The derived clump velocity dispersions ($\sigma_{\rm cl}$) range from $\sim$ 300 -- 600\,km\,s$^{-1}$, which correspond to a dynamical halo mass of $M_{\rm dyn} \sim 10^{13} M_{\odot}$, consistent with the predicted halo masses from the Millennium simulations (see Eq. 3 in \citealt{Li21} and discussions therein). The clump outflow velocities ($v_{\rm cl}$) range from $\sim$ 100 -- 250\,km\,s$^{-1}$, which yield considerable cloud lifetimes (or even cloud growth) in the hot ICM \citep{Gronke18,Li20,Li21}. Future observations may provide additional constraints on the relative velocities between the cool clumps and the ICM derived from simplistic configurations in this work. 

In terms of the survival of cool clumps, it is also helpful to consider the thermal sound speed of the hot ICM, $c_{s,\,{\rm ICM}} \sim \sqrt{k_{B}\,T_{\rm ICM}/m_{\rm p}} \sim 100 \rm\,km\,s^{-1}$. Therefore, the Mach numbers of the cool clumps $\mathcal{M}_{\rm cl} \equiv v_{\rm cl}/c_{s,\,\rm ICM}$ or $\sigma_{\rm cl}/c_{s,\,\rm ICM}$ $ \sim 1 - 2.5$, which are transonic or mildly supersonic. These Mach numbers are realistic for circumgalactic gas, and may slightly affect the dynamics of the cool clumps in the hot medium (see e.g. \citealt{Scannapieco15, Sparre20}).

\item \emph{Energy contribution from the inflowing gas}: We have observed signatures of gas inflow at the blob outskirts (see the blue-peak dominated spectra 3, 4, 14 and 15 and the inferred negative `outflow' velocities), and it is possible to estimate the associated cooling luminosity. As the blue-dominated profiles encompass most of the blob (see Figure \ref{fig:ratio_SB}), we assume that the clumps inflow in a semi-isotropic manner. The mass flow rate of the inflowing \HI\ gas is given by:
\begin{align}
\label{eqn:MHI} \dot{M_{\rm HI}} = 4\pi R_{\rm h}^2\,\rho_{\rm HI} \frac{dr}{dt} = 4\pi R_{\rm h}^2\,\rho_{\rm HI} v_{\rm inflow}
\end{align}
where $R_{\rm h}$ is the halo radius and $v_{\rm inflow}$ is the gas inflow velocity. The \HI\ mass density, $\rho_{\rm HI}$, is given by:
\begin{align}
\label{eqn:rho} \rho_{\rm HI} = \frac{M_{\rm HI}}{V_{\rm h}} = \frac{n_{\rm HI,\,cl}m_{\rm H}r_{\rm cl}^{3}N_{\rm cl}}{R_{\rm h}^{3}}
\end{align}
where $n_{\rm HI,\,cl}$ is the clump \HI\ number density, $m_{\rm H}$ is the mass of a hydrogen atom, and $r_{\rm cl}$ is the clump radius. The total number of clumps in the halo, $N_{\rm cl}$, is related to the volume filling factor:
\begin{align}
\label{eqn:nclumps} F_{\rm V} = \frac{N_{\rm cl}r_{\rm cl}^{3}}{R_{\rm h}^{3}}
\end{align}
With all these relations, the cooling luminosity generated from the released gravitational energy by gas infalling is:
\begin{gather*}
\label{eqn:Lcool}  L_{\rm cool} = \frac{GM_{\rm h}\dot{M_{\rm HI}}}{R_{\rm h}} = 4 \pi GM_{\rm h} F_{\rm V} f N_{\rm HI,\,cl} m_{\rm H} v_{\rm inflow} \\ 
= 1.4 \times 10^{42} F_{\rm V} \left(\frac{N_{\rm HI,\,cl}}{10^{17} {\rm cm^{-2}}} \right) \left(\frac{M_{\rm h}}{10^{13}M_{\odot}} \right) \left(\frac{v_{\rm inflow}}{100\,{\rm km\,s}^{-1}} \right) \rm erg\,s^{-1}
\end{gather*}
Note that $f$ is the radius ratio of the halo and the clumps (100 in our radiative transfer calculations). Taking $N_{\rm HI,\,cl} \sim 10^{18}\,\rm cm^{-2}$, and $F_{\rm V} \sim 0.3$, $M_{\rm h} \sim 10^{13} M_{\odot}$ and $v_{\rm inflow} \sim 100\,\rm km\,s^{-1}$(the mean of the derived inflow velocities), we get $L_{\rm cool} \sim 4\times10^{42}\,\rm erg\,s^{-1}$. This value is still more than one order of magnitude lower than the observed \lya\ luminosity of LAB2, even if all the cooling luminosity is emitted in \lya. Therefore, we conclude that the infalling of cool gas (cold accretion) plays a minor role in powering LAB2. This conclusion is consistent with the recent result of \citet{Ao20}, where they also found for a $z \sim$ 2.3 LAB that cool gas infalling helps produce blue-peak dominated \lya\ profiles, but is a subdominant powering mechanism compared to the photo-ionization process by embedded star-forming galaxies and/or AGNs\footnote{We have assumed that all the observed \lya\ emission comes from either photoionization by star-forming galaxies and/or AGN or accretion of infalling gas, because we were unable to identify any features that are suggestive of other powering mechanisms (e.g. shocks). The required SFR of ionizing sources inferred from the observed \lya\ luminosity is $\sim$ 80 M$_{\odot}$ / yr \citep{Kennicutt94}.}. 

\item \emph{Residual H\,{\textsc {i}} density in the ICM}: The derived \HI\ number densities correspond to column densities of $n_{\rm HI,\,{\rm ICM}}r_{\rm gal} \sim 10^{\rm 15} - 10^{\rm 18}\,\rm cm^{-2}$ (recall that $r_{\rm gal}$ = 5\,kpc is the radius of the simulation sphere) and \lya\ optical depths at line center of $\tau_{\rm 0, ICM} \sim 10 - 10^{\rm 3}$. Such column densities are high enough to contribute to the broadening of \lya\ spectra and produce (unsaturated) absorption at the line center. Taking the LAB2 halo radius as $R_{\rm h} \sim$ 50 kpc, the derived $n_{\rm HI,\,{\rm ICM}}$ values correspond to actual residual \HI\ number densities of $\sim 10^{\rm -8} - 10^{\rm -5}\,\rm cm^{-3}$. These values are moderately higher than the expected values assuming collisional ionization equilibrium, i.e. hydrogen number density $n_{\rm H,\,{\rm ICM}} \sim 10^{\rm -3} - 10^{\rm -2}\,\rm cm^{-3}$ and \HI\ fraction $x_{\rm HI,\,{\rm ICM}} \sim 10^{\rm -6}$ at $T_{\rm ICM}$ = 10$^{6}$K \citep{Dopita03}, but the difference is not significant especially considering that the highest $n_{\rm HI,\,{\rm ICM}}$ values (e.g. spectra 2 -- 5) appear close to the X-ray/submm continuum sources, which may be due to galactic feedback (see e.g. \S2.2 of \citealt{McQuinn16}).
\end{enumerate}

\subsubsection{Fitting Spectra at Different Spatial Positions Simultaneously}\label{sec:simultaneous_fit}

\begin{figure*}
\centering
\includegraphics[width=0.99\textwidth]{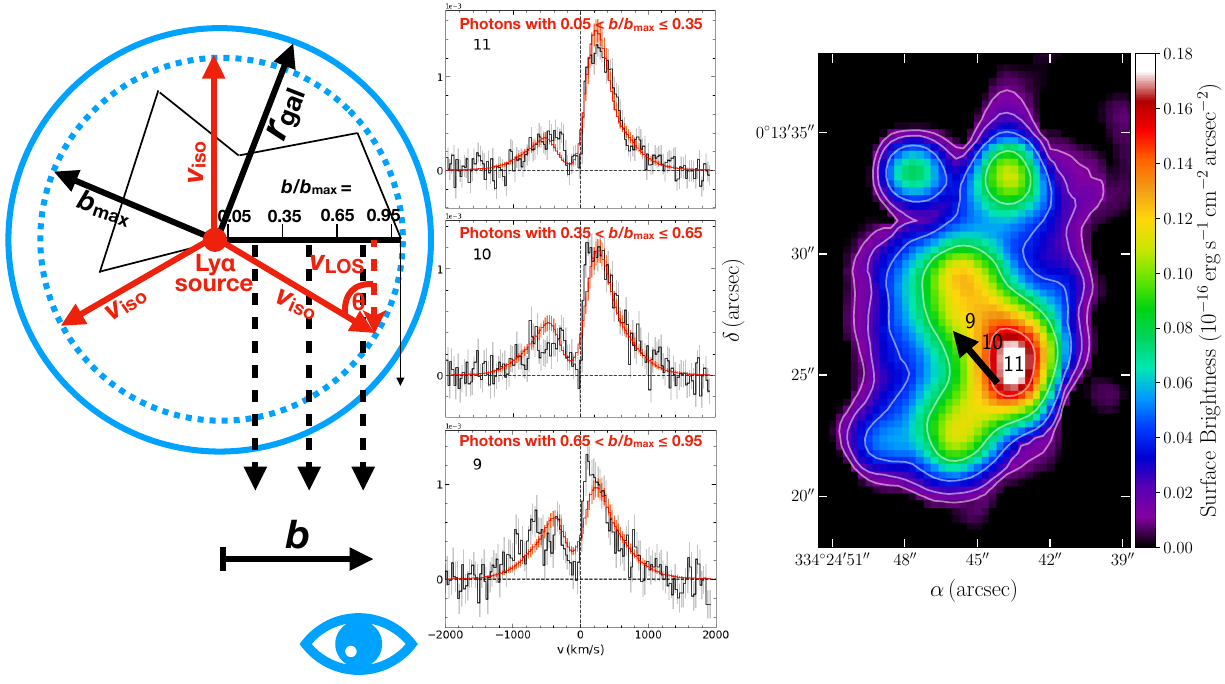}
    \caption{Results of fitting spatially-resolved \lya\ spectra (9, 10 and 11) at different impact parameters simultaneously. \emph{Left:} Illustration of how different photon bins are constructed in the multiphase, clumpy model. The ranges of impact parameters of three different \lya\ photon bins are: 0.05 $< b/b_{\rm max}$ $\leq$ 0.35, 0.35 $< b/b_{\rm max}$ $\leq$ 0.65 and 0.65 $< b/b_{\rm max}$ $\leq$ 0.95, where $b_{\rm max}$ is the largest impact parameter of all the scattered \lya\ photons. The solid red arrows represent an isotropic outflow of the cool clumps with velocity $v_{\rm iso}$, and the dashed red arrow represents the observed clump outflow velocity projected along the line-of-sight, $v_{\rm LOS}$. \emph{Middle:} Three aligned and equally-spaced \lya\ spectra (black, with grey 1-$\sigma$ error bars) and the corresponding binned best-fits (red, with orange 1-$\sigma$ Poisson errors). The likelihoods of the binned fits are comparable to those of the individual fits. \emph{Right:} The \lya\ SB map, with the alignment of three modeled spectra indicated by a black arrow.
    \label{fig:impact}}
\end{figure*}

In this section, we attempt to fit multiple spatially-resolved \lya\ profiles with multiphase, clumpy models in a more self-consistent way. We demonstrate that we can fit \lya\ spectra at different distances from a high SB center (i.e. at different impact parameters) simultaneously in one fitting run\footnote{We have also attempted to fit spectra at different spatial positions assuming an identified continuum source (e.g. M14 or the X-ray source) as the central \lya\ emitting source, but these attempts turned out to be unsuccessful. This is because in an outflow model the inner photons tend to be more red-dominated than the outer photons, which is opposite to the observed trend away from any continuum source (cf. \S\ref{sec:profiles})}. 

As illustrated in Figure \ref{fig:impact}, we chose spectra from three equally-spaced regions along a line (in projection) with respect to the high SB center (spectra 9, 10 and 11\footnote{These three spectra are located at $b/b_{\rm max}$ = 0 (i.e. down the barrel), 0.3 and 0.6, respectively (see \S\ref{sec:projection}).}) to perform our fit. For each model calculated in the fitting procedure, the \lya\ photon packages (10000 in total) are separated into three different bins, according to the impact parameter\footnote{Here we define impact parameter $b$ as the projected distance from the highest SB center relative to the line-of-sight. In the model, the impact parameter is determined as the distance from the center of the simulated region to the trajectory of the photon after the last scattering.} of their last-scattering locations: 0.05 $< b/b_{\rm max}$ $\leq$ 0.35, 0.35 $< b/b_{\rm max}$ $\leq$ 0.65 and 0.65 $< b/b_{\rm max}$ $\leq$ 0.95, where $b_{\rm max}$ is the largest impact parameter of all the scattered \lya\ photons. In this way, the representative photons included in each bin are numerous enough to construct a meaningful model spectrum. At every likelihood evaluation call, each observed \lya\ spectrum is compared to the corresponding `binned' model spectrum, and the likelihood is simply the sum of the likelihoods of three binned models.

The fitting results are shown in Figure \ref{fig:impact}. We found that the likelihoods of the binned fits are comparable to those of the individual fits (which were obtained by fitting spectra at each location independently). Moreover, the derived $v_{\rm cl}$ of this `combined fit' is fully consistent with the value of the individual fit of spectrum 11. One notable discrepancy is that the outermost binned model (0.65 $< b/b_{\rm max}$ $\leq$ 0.95) failed to fully reproduce the \lya\ flux density minimum near the systemic velocity of spectrum 9, which may be due to the assumption that $n_{\rm HI,\,{\rm ICM}}$ is constant over the whole simulated region, whereas in reality the residual \HI\ density in the ICM may vary spatially or extend beyond the multiphase medium. It is therefore possible to remedy this mismatch by considering models with radially-varying $n_{\rm HI,\,{\rm ICM}}$. Nevertheless, we have demonstrated the possibility of fitting spectra with different impact parameters simultaneously in a self-consistent manner. Further applications and extensions to the model are left to future work.

\subsubsection{v$_{\rm cl}$ v.s. b: A Simple Projection?}\label{sec:projection}
As presented in Table \ref{tab:Fitting_results}, the inferred outflow velocities of the cool clumps, $v_{\rm cl}$, vary significantly at different positions. Here we test whether this variation can be consistent with being a projection effect, i.e. the derived $v_{\rm cl}$ values from the multiphase, clumpy models are simply the projected line-of-sight components of an isotropic outflow velocity.

\begin{table}
\renewcommand\arraystretch{1.8}
  \scriptsize \caption{Best-fit parameters of fitting spectra 9, 10 and 11 simultaneously with the multiphase, clumpy model.}
  \label{tab:Fitting_results_combined}
  \centering
  \begin{adjustbox}{angle=0}
  \setlength{\tabcolsep}{3pt}
  \begin{tabular}{cccccc}%
\hline%
\hline%
${\rm log}\,n_{\rm HI,\,{\rm ICM}}$&$F_{\rm V}$&${\rm log}\,N_{\rm HI,\,{\rm cl}}$&$\sigma_{\rm cl}$&$v_{\rm cl}$&$\Delta v_{\rm clumpy}$\\%
(cm$^{-3})$&&(cm$^{-2})$&(km\,s$^{-1})$&(km\,s$^{-1})$&(km\,s$^{-1})$\\%
(1)&(2)&(3)&(4)&(5)&(6)\\%
\hline%
-6.0$^{+0.2}_{-0.2}$&0.07$^{+0.01}_{-0.01}$&18.2$^{+0.1}_{-0.1}$&463$^{+78}_{-33}$&193$^{+24}_{-15}$&176$^{+10}_{-10}$\\%

\hline%
\hline%
\end{tabular}
  \end{adjustbox}
  \begin{tablenotes}
  \small
    \item \textbf{Notes.} Results of fitting spectra 9, 10 and 11 simultaneously with the multiphase, clumpy model, as described in \S\ref{sec:simultaneous_fit}. The columns are: (1) the \HI\ number density in the ICM; (2) the cloud volume filling factor; (3) the clump \HI\ column density; (4) the velocity dispersion of the clumps; (5) the radial outflow velocity of the clumps; (6) the velocity shift relative to the initial guess for the systemic redshift ($z$ = 3.098). The fitted parameters are given as averages and 16\% -- 84\% quantiles, i.e. 1-$\sigma$ confidence intervals.
   \end{tablenotes}
\end{table}

\begin{figure}
\includegraphics[width=0.45\textwidth]{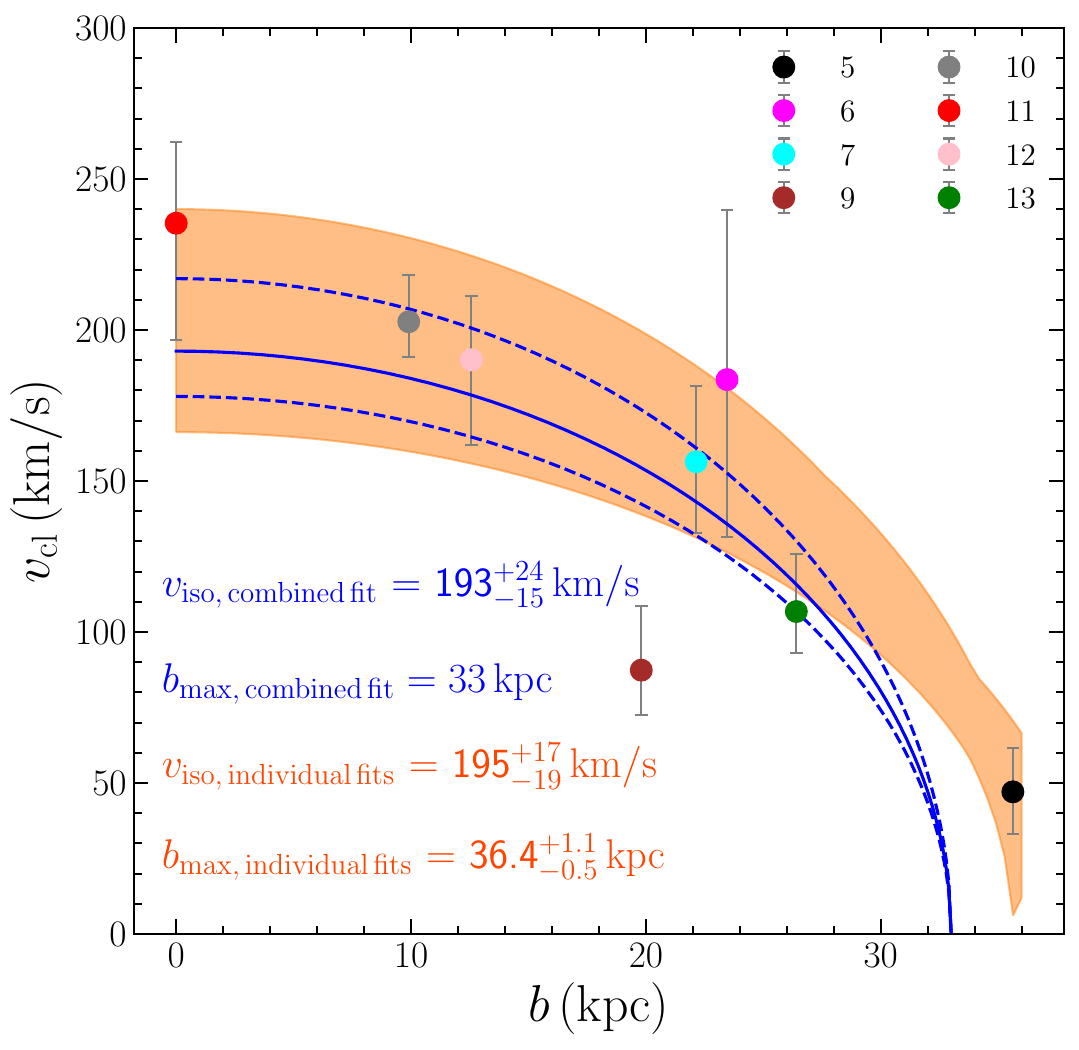}
    \caption{The relation between the clump outflow velocity ($v_{\rm cl}$) and the impact parameter ($b$). The color-coded points represent $v_{\rm cl}$ values (with 1-$\sigma$ error bars shown in grey) derived from individual fits of eight different \lya\ spectra from the south high SB region, and the orange shaded region represent the range of twenty best-fits (using Eq. \ref{eqn:vcl}) of the data points perturbed by their 1-$\sigma$ uncertainties. The best-fit parameters with 1-$\sigma$ uncertainties are also shown in the lower left. The blue solid and dashed curves are reference $v_{\rm cl}$ v.s. $b$ curves generated by setting $v_{\rm iso}$ = 193$_{-15}^{+24}$\,km\,s$^{-1}$ (derived from the combined fit) and $b_{\rm max}$ = 33\,kpc (calculated by mapping the geometric distances in the model to the actual physical distances in Eq. \ref{eqn:vcl}). The reference $v_{\rm cl}$ v.s. $b$ curve from the combined fit is fairly consistent with the $v_{\rm cl}$ values derived from individual fits (as well as the fitted $v_{\rm iso}$ and $b_{\rm max}$ values) within 1-$\sigma$ uncertainties, suggesting the variations in $v_{\rm cl}$ may be simply the projection of a radial outflow along the line-of-sight. 
    \label{fig:vcl_b}}
\end{figure}

The line-of-sight component of an isotropic outflow velocity is given by (as illustrated in the left panel of Figure \ref{fig:vcl_b}):

\begin{align}
\label{eqn:vcl} v_{\rm LOS} &= v_{\rm iso} {\rm cos} \theta \nonumber \\
& = v_{\rm iso} \sqrt{1-(b/b_{\rm max})^2}
\end{align}
where $v_{\rm LOS}$ is the observed line-of-sight outflow velocity, $v_{\rm iso}$ is the isotropic outflow velocity, $\theta$ is the angle between the line-of-sight and the isotropic outflow velocity, $b$ is the impact parameter of the observed location, and $b_{\rm max}$ is the maximum impact parameter of the observed region (or the radius if the region is spherically symmetric). We further examined Eq. (\ref{eqn:vcl}) by plotting the $v_{\rm cl}$ values derived from individual fits\footnote{Here we selected eight red-peak dominated spectra that are close to the \lya\ SB center (i.e. spectrum 11) from the south high SB region: spectra 5, 6, 7, 9, 10, 11, 12, 13.} along with the range of fitted $v_{\rm cl}$ v.s. $b$ curves in the right panel of Figure \ref{fig:vcl_b}. A reference $v_{\rm cl}$ v.s. $b$ curve from the combined fit of spectra 9, 10 and 11 is also plotted for comparison, which is generated by setting $v_{\rm iso}$ = 193$_{-15}^{+24}$\,km\,s$^{-1}$ (the $v_{\rm cl}$ of the combined fit of spectra 9, 10 and 11) and $b_{\rm max}$ = 33\,kpc (calculated by mapping the geometric distances in the model to the actual physical distances\footnote{Specifically, the geometric distance between spectra 9 and 10 in the model is 0.3\,$b_{\rm max}$, and the actual physical distance between these two locations are 9.9\,kpc. Hence $b_{\rm max}$ = 33\,kpc.}) in Eq. (\ref{eqn:vcl}). 

As can be seen in Figure \ref{fig:vcl_b}, the reference $v_{\rm cl}$ v.s. $b$ curve from the combined fit (blue solid and dashed lines) is fairly consistent with the $v_{\rm cl}$ values derived from individual fits (as well as the fitted $v_{\rm iso}$ and $b_{\rm max}$ values) within 1-$\sigma$ uncertainties, suggesting the variations in $v_{\rm cl}$ may be simply the projection of a radial outflow along the line-of-sight. Therefore, we conclude that the observed variation in $v_{\rm cl}$ with respect to $b$ can be reasonably accounted for by a simple line-of-sight projection effect. This would also naturally explain the observed increase in $F_{\rm blue}/F_{\rm red}$ ratio towards the blob outskirts, as $F_{\rm blue}/F_{\rm red}$ is inversely correlated with $v_{\rm cl}$ (see \S\ref{sec:profiles}), although the blue-dominated spectra at the very largest distances require inflows.

\section{Previous Studies of LAB2}\label{sec:previous}
Here we summarize the results of previous studies of LAB2 and compare with the present work. We classify different studies according to the wavelengths of their observations:

\textbf{Optical}: \citet{Steidel2000} first discovered LAB2 using narrow-band imaging and provided its angular size ($\sim$\,15$\arcsec$), \lya\ luminosity ($\sim\,9.0 \times 10^{43}$\,erg\,$\rm s^{-1}$) and limits on the rest-frame equivalent width ($\gtrsim$\,370\,\AA). They also identified a velocity shear of $\sim$\,2000\,km s$^{-1}$ with respect to the LBG, M14. 

\citet{Wilman05} carried out IFU observations using SAURON on the 4.2 m William Herschel Telescope (WHT) on La Palma. They observed the ubiquitous `double-peak + central trough' feature of most \lya\ profiles in LAB2, and modeled the profiles with a Gaussian emission line and a superimposed Voigt profile absorber. They claimed that the profiles can be explained by an `intrinsic \lya\ emission + \HI\ shell absorption' model, where the shell is the cool material swept-up by starburst-driven outflows (`superwinds'). They also noted that a large foreground absorber that covers the whole blob is unlikely, as the absorber would have an unreasonably large size.

\citet{Martin14} made further IFU observations using the Palomar Cosmic Web Imager (PCWI). They proposed that the \lya\ emission could be produced by either AGN fluorescence or gravitational cooling radiation. They also claimed that there is evidence of both inflows and outflows at different viewing angles, which is consistent with our findings that both blue-dominated and red-dominated \lya\ profiles exist in LAB2.

\textbf{X-ray}: \citet{Basu04} first detected an obscured hard X-ray source in the \emph{Chandra} 2 -- 8 keV band. They claimed that the unabsorbed X-ray luminosity ($\sim 10^{44}$\,erg\,$\rm s^{-1}$) is consistent with an AGN. Deeper \emph{Chandra} observations have been carried out in \citet{Lehmer09cat, Lehmer09AGN}, based on which \citet{Geach09} claimed that the UV luminosity of the AGN alone is sufficient to power the whole blob via photoionization. However, as we have shown in Figure \ref{fig:Lya_images} and subsequent analyses, the location of the AGN with respect to the \lya\ emission is difficult to reconcile with the \lya\ morphology.

\textbf{IR}: \citet{Geach07} carried out \emph{Spitzer} observations at IRAC (3.6 – 8 $\mu$m) and MIPS (24 $\mu$m). They reported detections at three positions in LAB2 (named a, b and c), where b is likely the counterpart of the X-ray source, and c is a foreground source at lower redshift\footnote{We found that detection c is close to the position where the \lya\ SB is the highest (see the left panel of Figure \ref{fig:Lya_images}). Unfortunately, its mid-IR colors suggest that it should be a foreground galaxy at $z$ < 1 \citep{Geach07, Webb09}. In our KCWI datacube, the location of its Ca II H and K lines suggests a redshift of $z$ = 0.213.}. \citet{Webb09} further confirmed the detection of a and b in the IRAC bands, and we have marked their positions in Figure \ref{fig:Lya_images}.

\textbf{Submm}: \citet{Chapman01} first reported a 3.3\,$\pm$\,1.2 mJy detection at 850 $\mu$m within the $\sim$15$\arcsec$ beam of SCUBA. \citet{Geach05} studied the relation between the \lya\ luminiosity and the bolometric luminosity of 23 LABs and noted that LAB2 is an outlier of the relation, which could be due to the effect of AGN and other environmental factors. However, a recent study by \citet{Hine16b} reported a non-detection at 850 $\mu$m using SCUBA-2. Another recent study by \citet{Ao17} reported a significant 0.91\,$\pm$\,0.10 mJy detection at 850 $\mu$m using ALMA, which coincides with the X-ray detection from \citet{Lehmer09cat}.

Previous studies on LAB2 focused mainly on qualitatively studying the continuum sources (e.g. the LBG M14 and the X-ray source) in terms of their energy budgets and/or analysing the observed \lya\ profiles with empirical tools without carrying out radiative transfer calculations. In contrast, the present work carefully examines the spatially-resolved \lya\ profiles (such as mapping their blue-to-red flux ratios, see \S\ref{sec:profiles}) and decodes them using radiative transfer calculations with multiphase, clumpy models (see \S\ref{sec:RT}). In our modelling, we assumed that a central powering source exists near the highest \lya\ SB regions, although it is still puzzling that there is no viable continuum source coincident with the \lya\ SB peak (see \S\ref{sec:oiii}). Not only have we successfully reproduced fifteen representative \lya\ profiles with realistic physical parameters, but we also managed to fit \lya\ profiles at different impact parameters consistently, and explained the observed spatial variation in the $F_{\rm blue}/F_{\rm red}$ ratio and outflow velocity. These results support the `central powering + scattering' scenario, i.e. the \lya\ photons are generated by central powering source(s) and then scatter with outflowing, multiphase \HI\ gas while propagating outwards. We have also observed signatures of accretion of infalling cool gas at the blob outskirts. As we have shown in \S\ref{sec:results}, although the infalling of cool gas is responsible for shaping the observed blue-dominated \lya\ profiles, its energy contribution is likely to be minor compared to the photo-ionization by central (as yet unidentified) sources.

\section{Conclusions}\label{sec:conclusion}
We present new deep spectroscopic observations of SSA22-LAB2 at $z$ = 3.1 using KCWI and MOSFIRE. The main conclusions of our analysis are:

\begin{enumerate}
\item By creating a narrow-band \lya\ image, we observed extended \lya\ emission in three distinct regions, among which the south region is the largest and has a high \lya\ SB center that is far away from known continuum sources;

\item We found that the \lya\ profiles are dominated by a red peak in regions of high \lya\ SB, but tend to be more symmetric and even blue-peak dominated in the low SB outskirts. The median blue-to-red flux ratio is anti-correlated with \lya\ SB, which may be due to the decrease of the projected line-of-sight outflow velocity in the periphery of the halo;

\item We searched through the two MOSFIRE slits that had been observed near to the highest \lya\ SB regions, and found no significant detection of nebular emission within the region of \lya\ emission;

\item To decode the spatially-resolved \lya\ profiles using Monte-Carlo radiative transfer (MCRT) modelling, we used both multiphase, clumpy models and shell models, both of which successfully reproduced the diverse \lya\ morphologies. We found a significant correlation between parameters of the two different models, and our derived parameters may alleviate the previously reported discrepancies between the shell model parameters and data;

\item We have managed to fit \lya\ spectra at different impact parameters simultaneously assuming a common central source. We also found that the variation of the clump outflow velocity with respect to impact parameter can be approximately explained as a simple line-of-sight projection effect of a radial outflow;

\item We conclude that our results support the `central powering +scattering' scenario, i.e. the \lya\ photons are generated by a central powering source and then scatter with outflowing, multiphase \HI\ gas while propagating outwards. The infalling of cool gas is responsible for shaping the observed blue-dominated \lya\ profiles, but the energy contribution of infalling material to the total \lya\ luminosity is less than 10\%, i.e. minor compared to the photo-ionization by star-forming galaxies and/or AGNs.
\end{enumerate}

\section*{Acknowledgements}

We thank the anonymous referee for carefully reading our manuscript and providing constructive feedback, which significantly improved the quality of this paper. We thank Phil Hopkins for providing computational resources. ZL, CCS and YC acknowledge financial support by NSF grant AST-2009278. YM acknowledges support from JSPS KAKENHI Grant (17H04831, 17KK0098, 19H00697 and 20H01953). The data presented herein were obtained at the W. M. Keck Observatory, which is operated as a scientific partnership among the California Institute of Technology, the University of California and the National Aeronautics and Space Administration. The Observatory was made possible by the generous financial support of the W. M. Keck Foundation. We are also grateful to the dedicated staff of the W.M. Keck Observatory who keep the instruments and telescopes running effectively. MG was supported by NASA through the NASA Hubble Fellowship grant HST-HF2-51409 awarded by the Space Telescope Science Institute, which is operated by the Association of Universities for Research in Astronomy, Inc., for NASA, under contract NAS5-26555. Numerical calculations were run on the Caltech compute cluster ``Wheeler,'' allocations from XSEDE TG-AST130039 and PRAC NSF.1713353 supported by the NSF, and NASA HEC SMD-16-7592. This research made use of Montage. It is funded by the National Science Foundation under Grant Number ACI-1440620, and was previously funded by the National Aeronautics and Space Administration's Earth Science Technology Office, Computation Technologies Project, under Cooperative Agreement Number NCC5-626 between NASA and the California Institute of Technology. We also acknowledge the use of the the following software packages: Astropy \citep{Astropy18}, the SciPy and NumPy system \citep{Scipy20, Numpy20}, seaborn \citep{Waskom2021} and QFitsView\footnote{https://www.mpe.mpg.de/ ott/QFitsView/}.

\section*{Data availability}
The data underlying this article will be shared on reasonable request to the corresponding author. 

\bibliographystyle{mnras}
\bibliography{LAB2}

\appendix

\section{Note on The Usage of RT Models and Interpretation of Fitting Results}\label{sec:justification}
In this work, we have used the multiphase, clumpy model to fit spatially-resolved \lya\ spectra. Although the multiphase, clumpy model adopted in this work assumes a central \lya\ source surrounded by an ensemble of gas clouds distributed isotropically, we emphasize that such an assumption does not affect our attempt to fit these models to the observed individual spatially-resolved spectra as an approximate way to extract the local gas properties -- e.g. ${\rm H\,{\textsc {i}}}$ column densities, clump velocity dispersions, and particularly, clump outflow velocities {\it{along the line-of-sight}} relative to the local systemic redshift.

To corroborate our point, we design an experiment where we run a set of four multiphase, clumpy models with different $v_{\rm cl}$ but otherwise the same parameters. We then plot the \lya\ spectra in the ``down the barrel'' spatial bin of the first model and cos$\theta$ (= $\sqrt{1-(b/b_{\rm max})^2}$, see Eq. \ref{eqn:vcl}) = $v_{\rm cl,\,1} / v_{{\rm cl},\,i}$ spatial bin of the $i$th model ($i$ = 2, 3, 4), so that the outflow velocities along the line-of-sight for all four models are the same. We show below that these four binned \lya\ spectra are consistent. 

\begin{figure}
\centering
\includegraphics[width=0.45\textwidth]{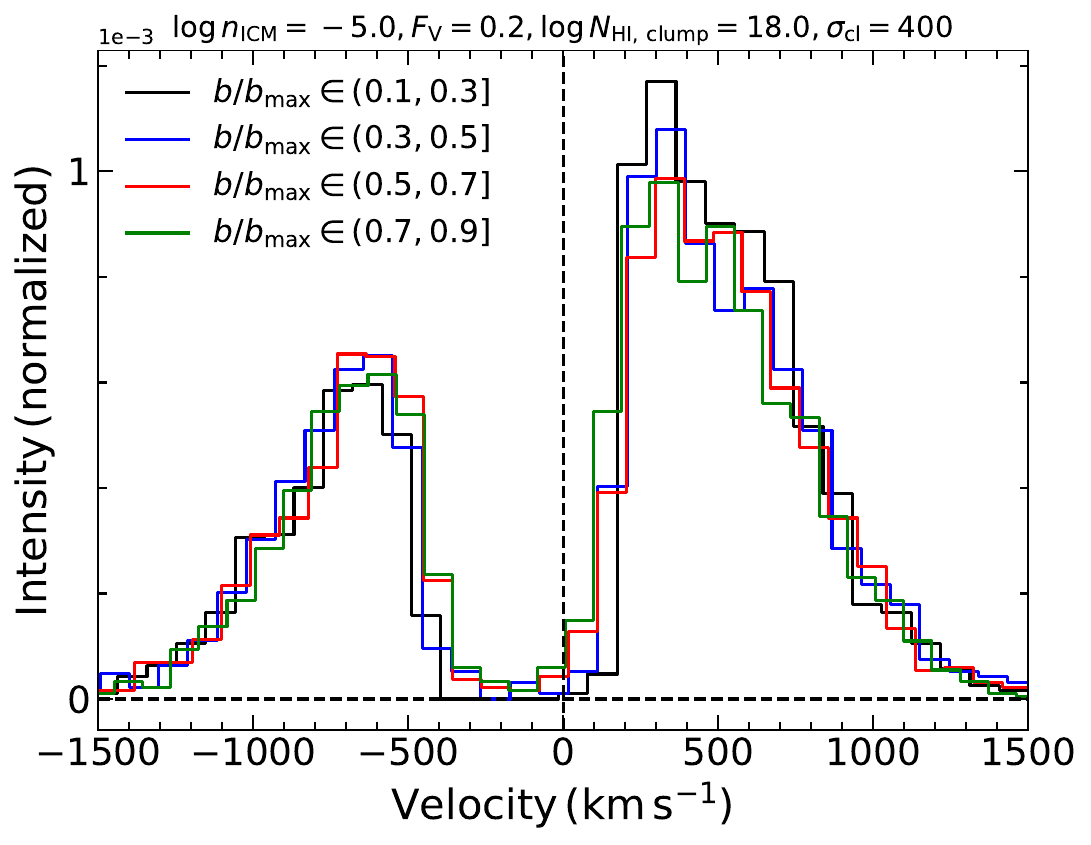}
    \caption{Justification for using the multiphase, clumpy models to fit the spatially-resolved Ly$\alpha$ profiles. four multiphase, clumpy models have (${\rm log}\,n_{\rm ICM}\,(\rm cm^{-3})$, $F_{\rm V}$, ${\rm log}\,N_{\rm HI,\,cl}\,(\rm cm^{-2})$, $\sigma_{\rm cl}\,(\rm km\,s^{-1})$) = (-5.0, 0.2, 18.0, 400) and $v_{\rm cl}\,(\rm km\,s^{-1})$ = (100, 107, 122, 163), respectively. The photons from the following ranges of impact parameters are selected to construct the \lya\ spectra for each of the four models: $b/b_{\rm max}\,\in $(0.1, 0.3], (0.3, 0.5], (0.5, 0.7], and (0.7, 0.9], respectively. The line-of-sight $v_{\rm cl}$ for the $b/b_{\rm max}\,\in $(0.1, 0.3] photons (i.e. nearly "down the barrel") of the first model and for the other three bins of photons (i.e. "off the barrel") are basically the same. The \lya\ model spectra constructed from these four bins of photons are consistent with each other within 1-$\sigma$ uncertainties (about 10\% assuming Poisson photon distribution).
    \label{fig:projection}}
\end{figure}

We hereby illustrate our point with a specific example (whose parameters are typical in our fitting) in Figure \ref{fig:projection}: four multiphase, clumpy models have (${\rm log}\,n_{\rm ICM}\,(\rm cm^{-3})$, $F_{\rm V}$, ${\rm log}\,N_{\rm HI,\,cl}\,(\rm cm^{-2})$, $\sigma_{\rm cl}\,(\rm km\,s^{-1})$) = (-5.0, 0.2, 18.0, 400) and $v_{\rm cl}\,(\rm km\,s^{-1})$ = (100, 107, 122, 163), respectively. The photons from the following ranges of impact parameters are selected to construct the \lya\ spectra for each of the four models: $b/b_{\rm max}\,\in $(0.1, 0.3], (0.3, 0.5], (0.5, 0.7], and (0.7, 0.9], respectively. The line-of-sight $v_{\rm cl}$ for the $b/b_{\rm max}\,\in $(0.1, 0.3] photons (i.e. nearly ``down the barrel'') of the first model and for the other three bins of photons (i.e. ``off the barrel'') are basically the same (using Eq. \ref{eqn:vcl}). It can be seen that the \lya\ model spectra constructed from these four different bins of photons are consistent. Therefore, it is reasonable to extract local gas properties by fitting \lya\ spectra observed away from continuum sources, as long as we interpret the output clump outflow velocity as a projected, line-of-sight velocity with respect to the local systemic redshift.

\section{Posterior Probability Distributions of the Multiphase, Clumpy model parameters}\label{sec:posterior}

Here we present the joint and marginal posterior probability distributions of the multiphase, clumpy model parameters for all fifteen representative \lya\ spectra derived from nested sampling. 

\begin{figure*}
\centering
\includegraphics[width=\textwidth]{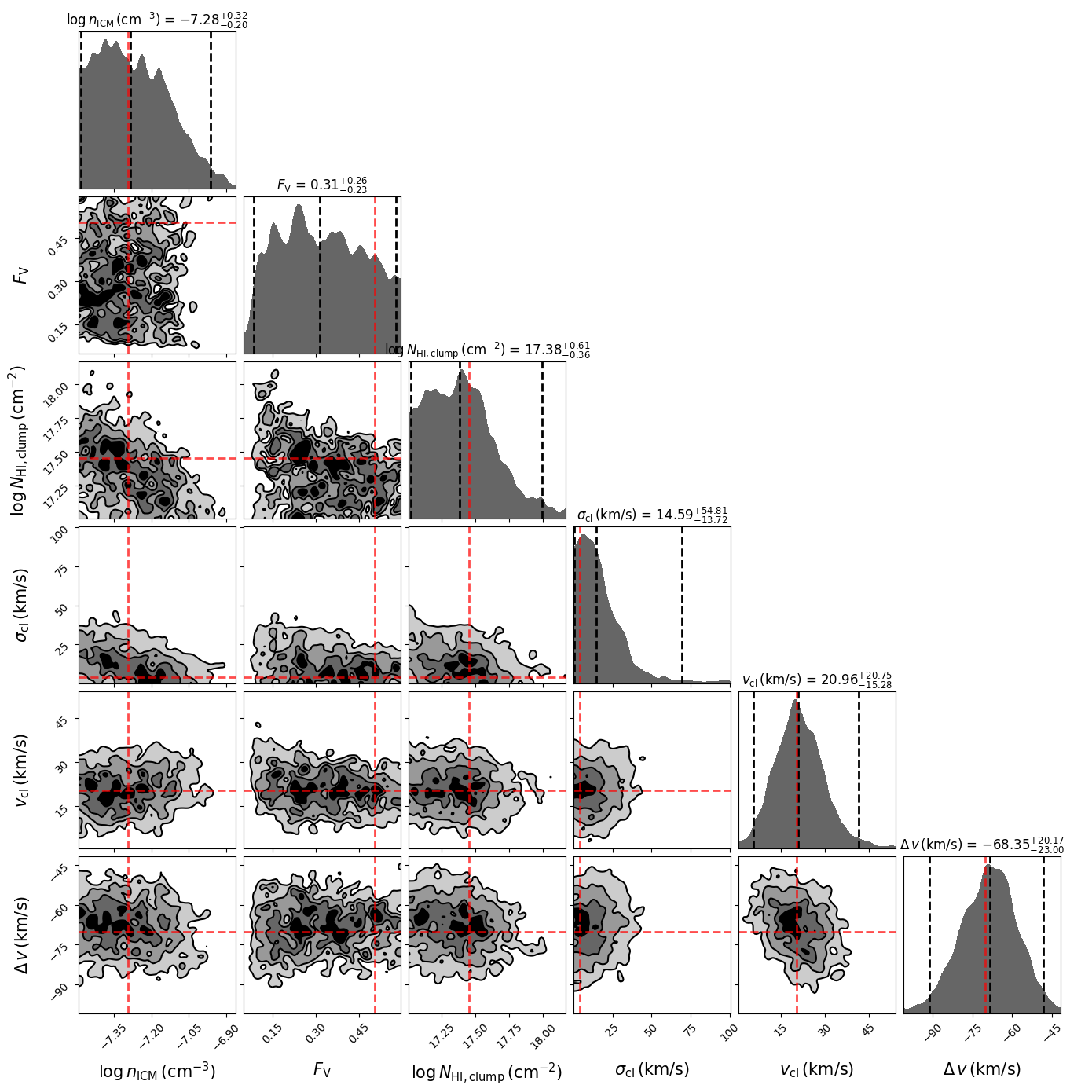}
\end{figure*}

\begin{figure*}
\centering
\includegraphics[width=\textwidth]{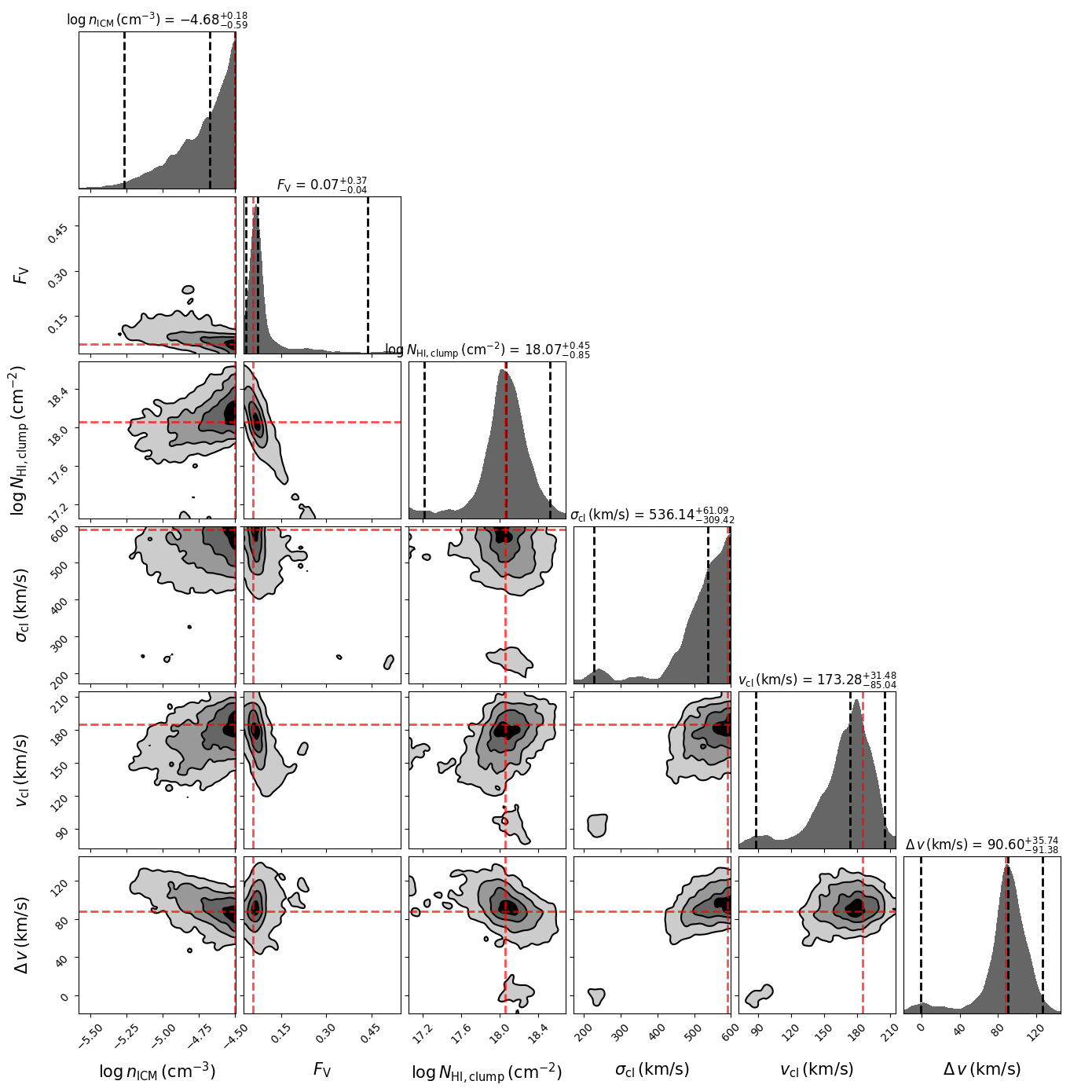}
\end{figure*}

\begin{figure*}
\centering
\includegraphics[width=\textwidth]{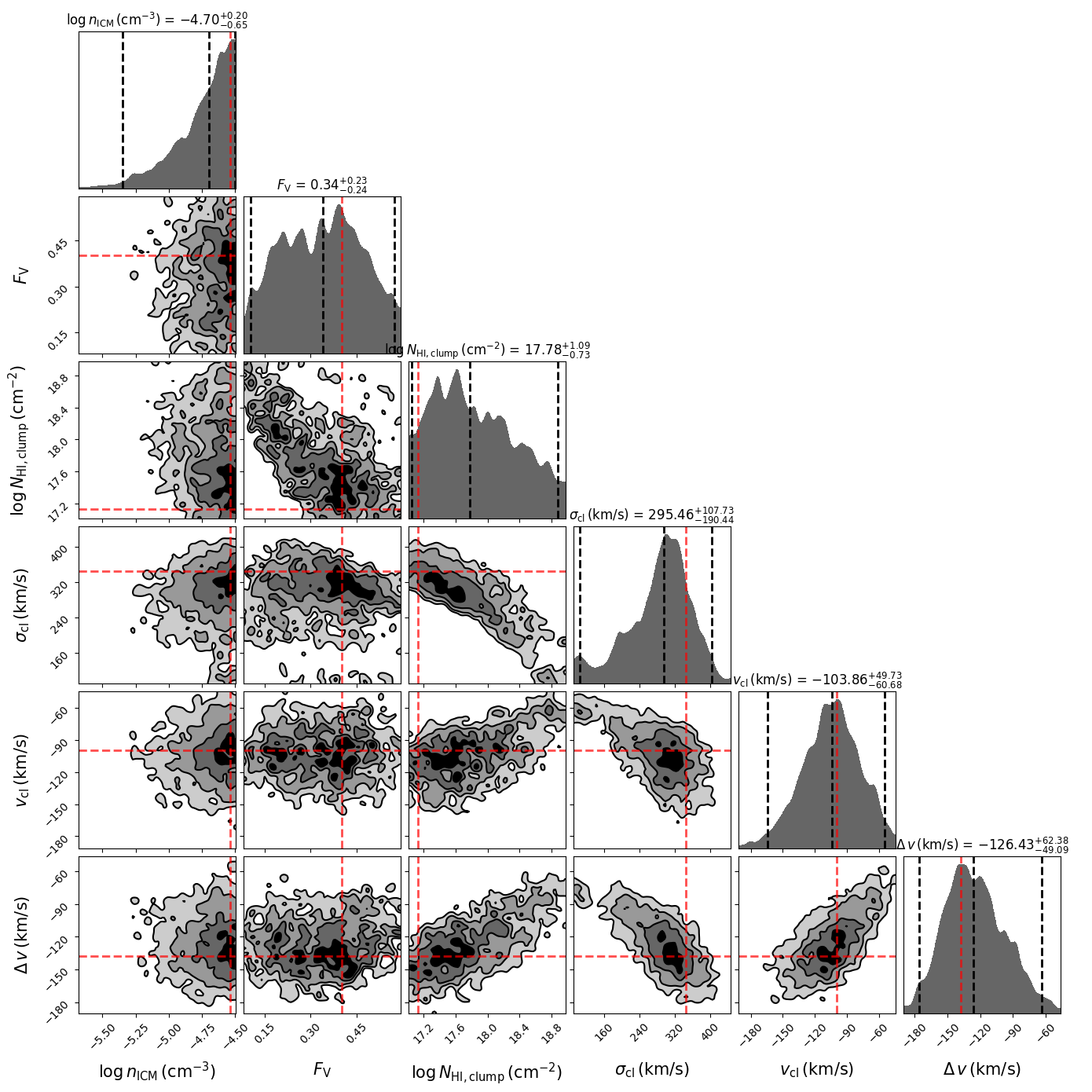}
\end{figure*}

\begin{figure*}
\centering
\includegraphics[width=\textwidth]{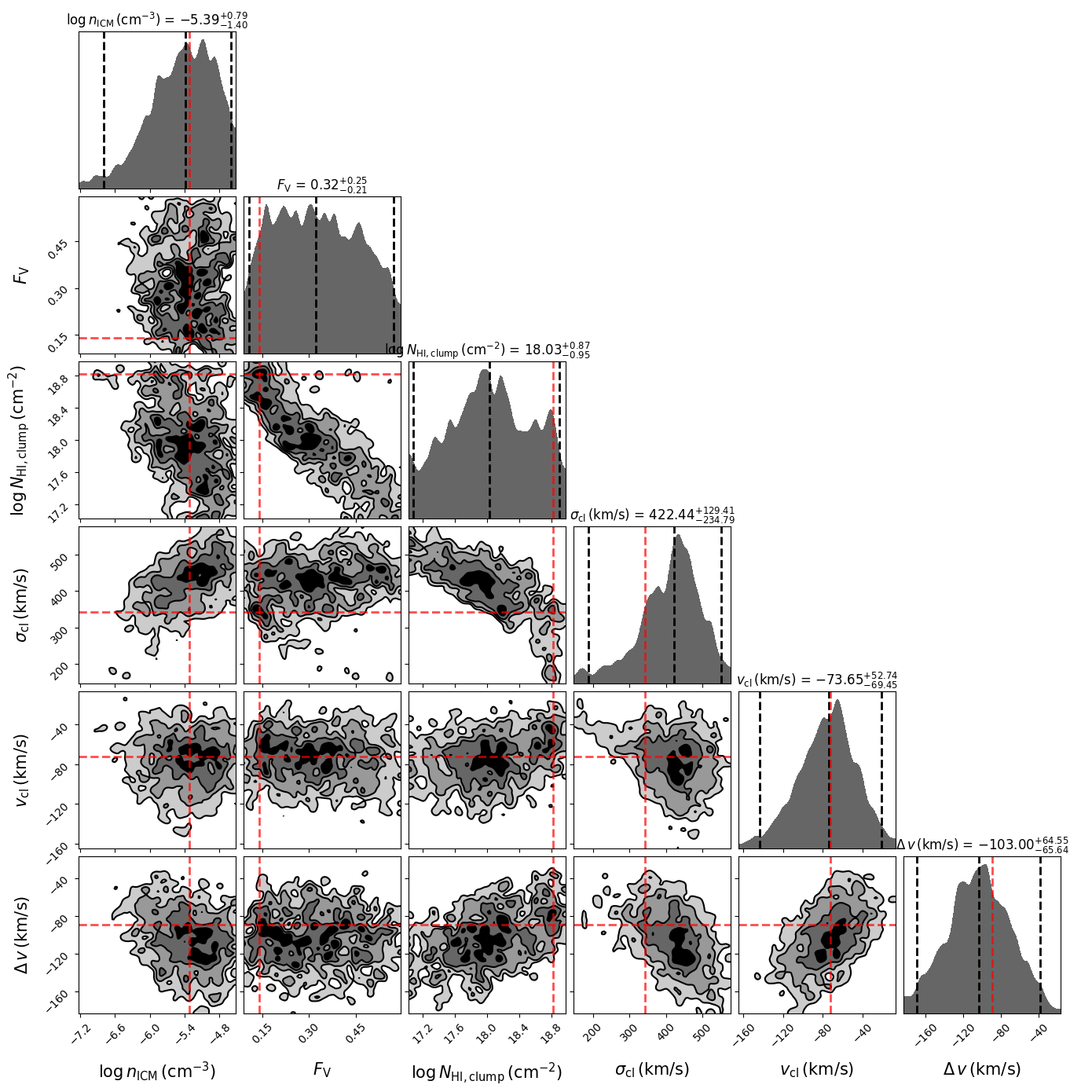}
\end{figure*}

\begin{figure*}
\centering
\includegraphics[width=\textwidth]{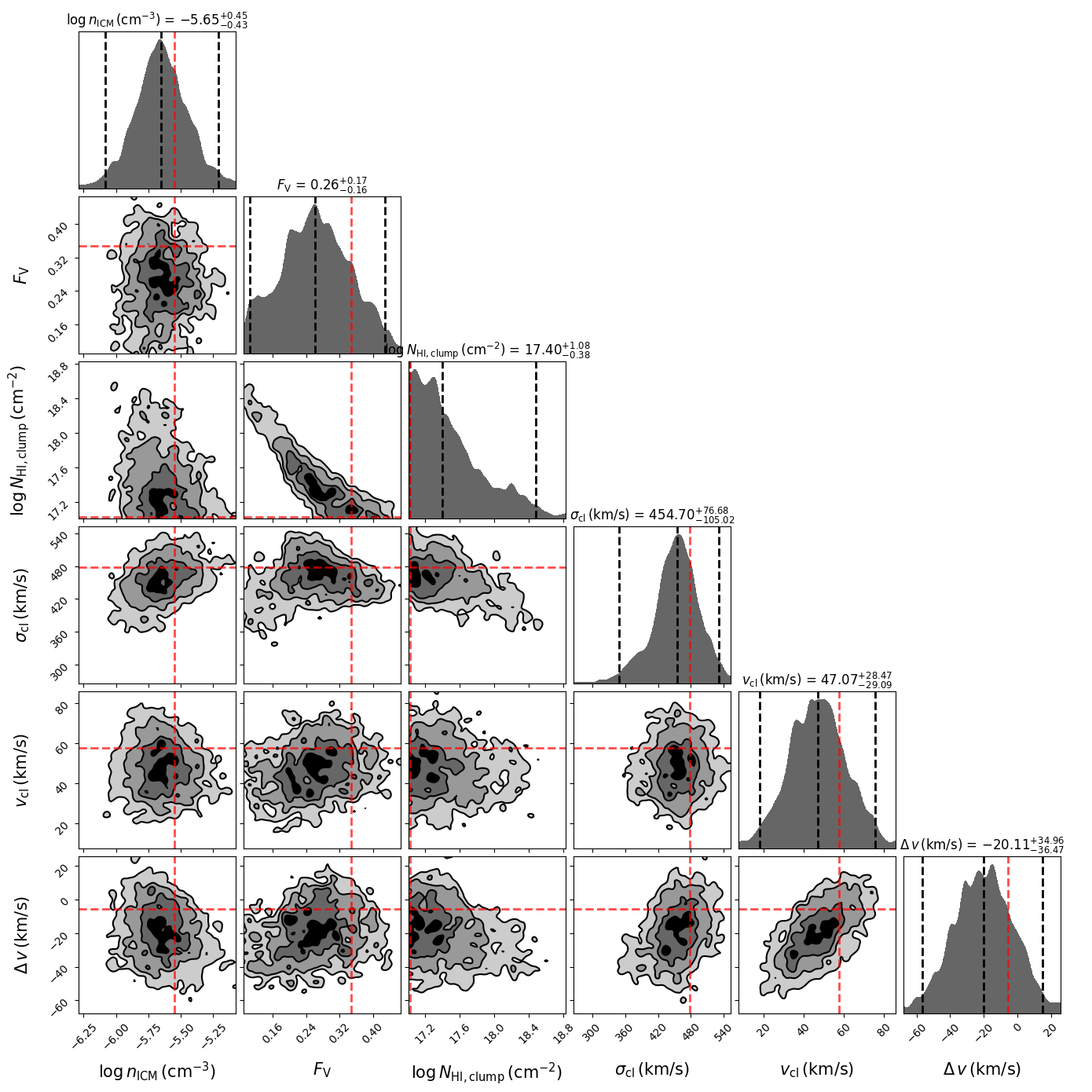}
\end{figure*}

\begin{figure*}
\centering
\includegraphics[width=\textwidth]{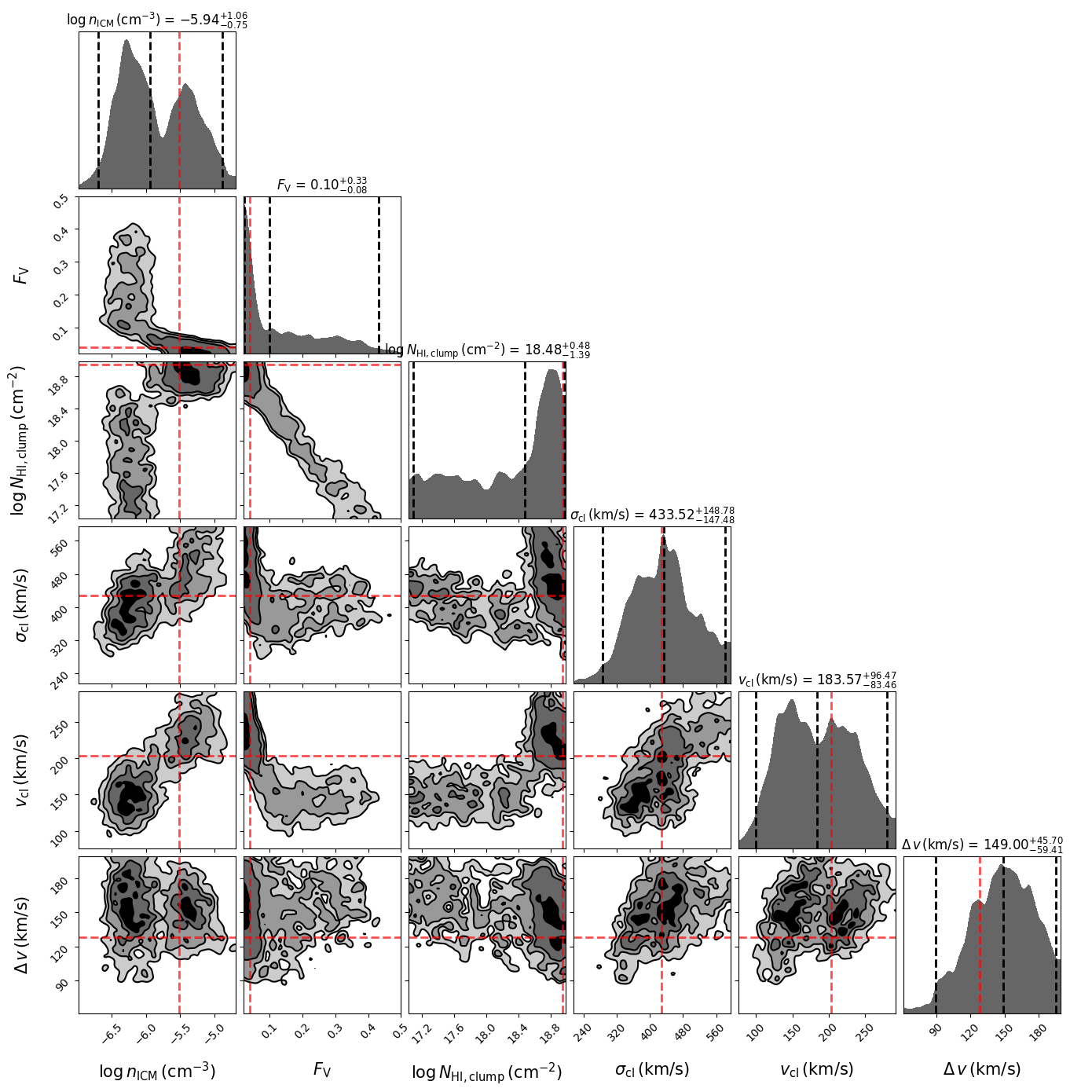}
\end{figure*}

\begin{figure*}
\centering
\includegraphics[width=\textwidth]{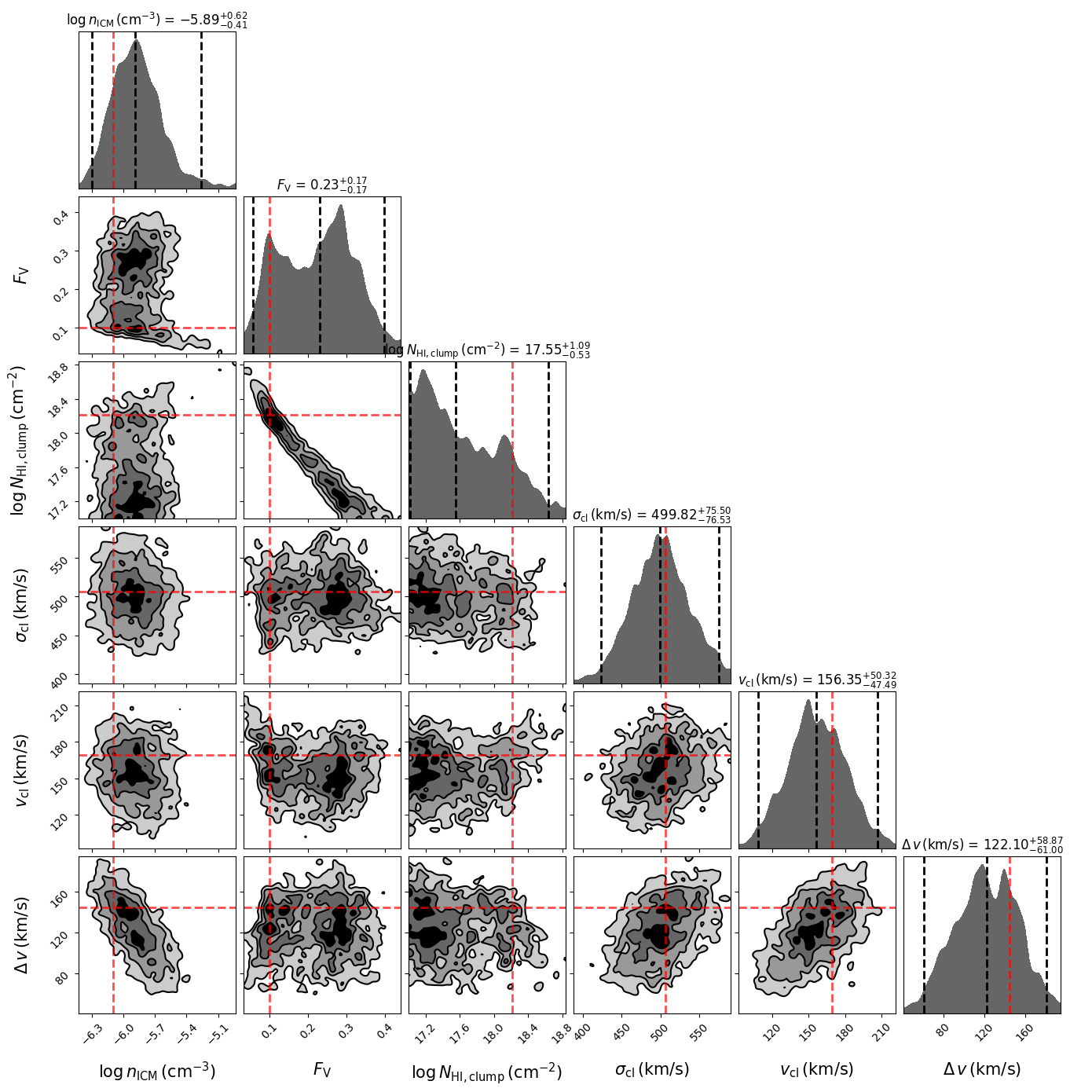}
\end{figure*}

\begin{figure*}
\centering
\includegraphics[width=\textwidth]{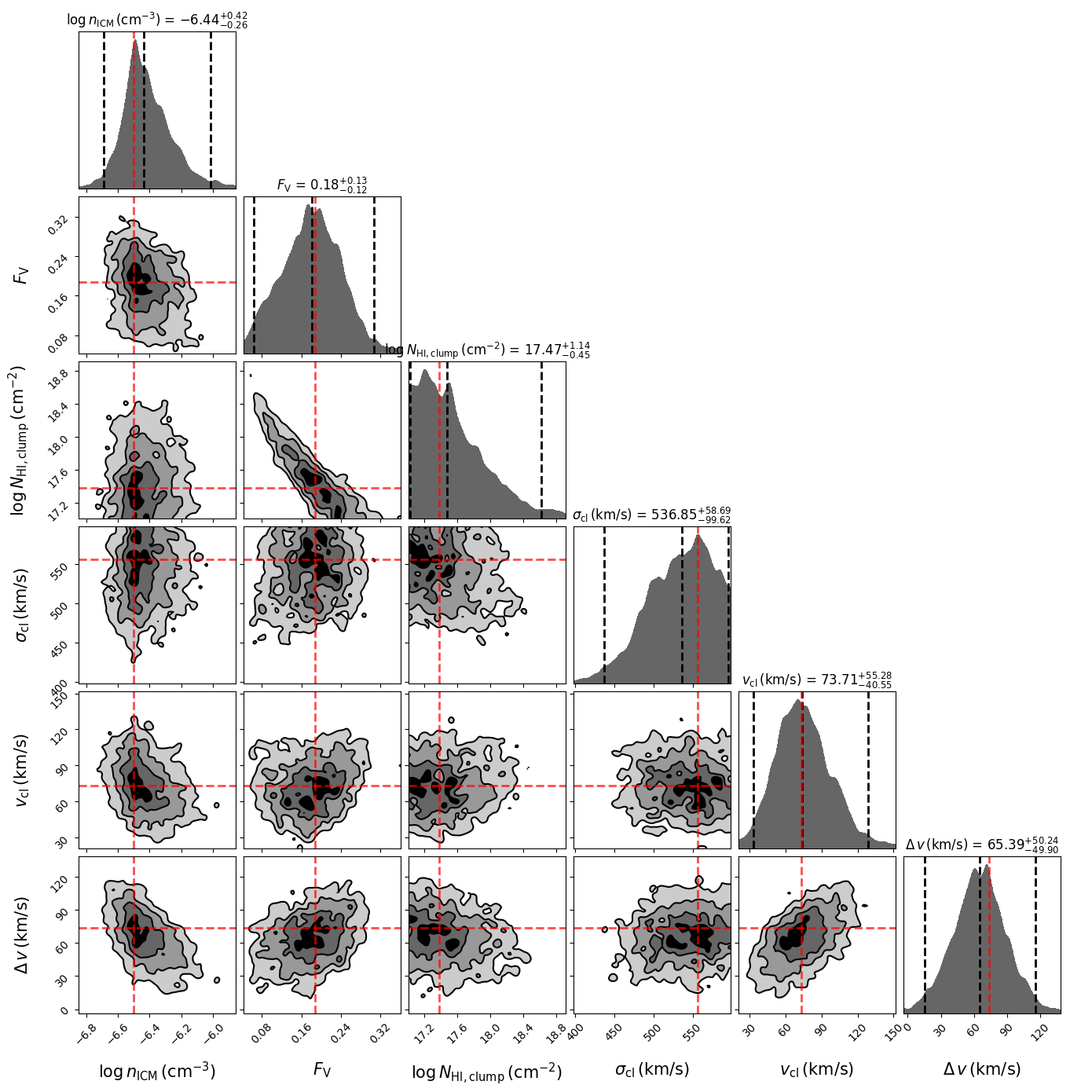}
\end{figure*}

\begin{figure*}
\centering
\includegraphics[width=\textwidth]{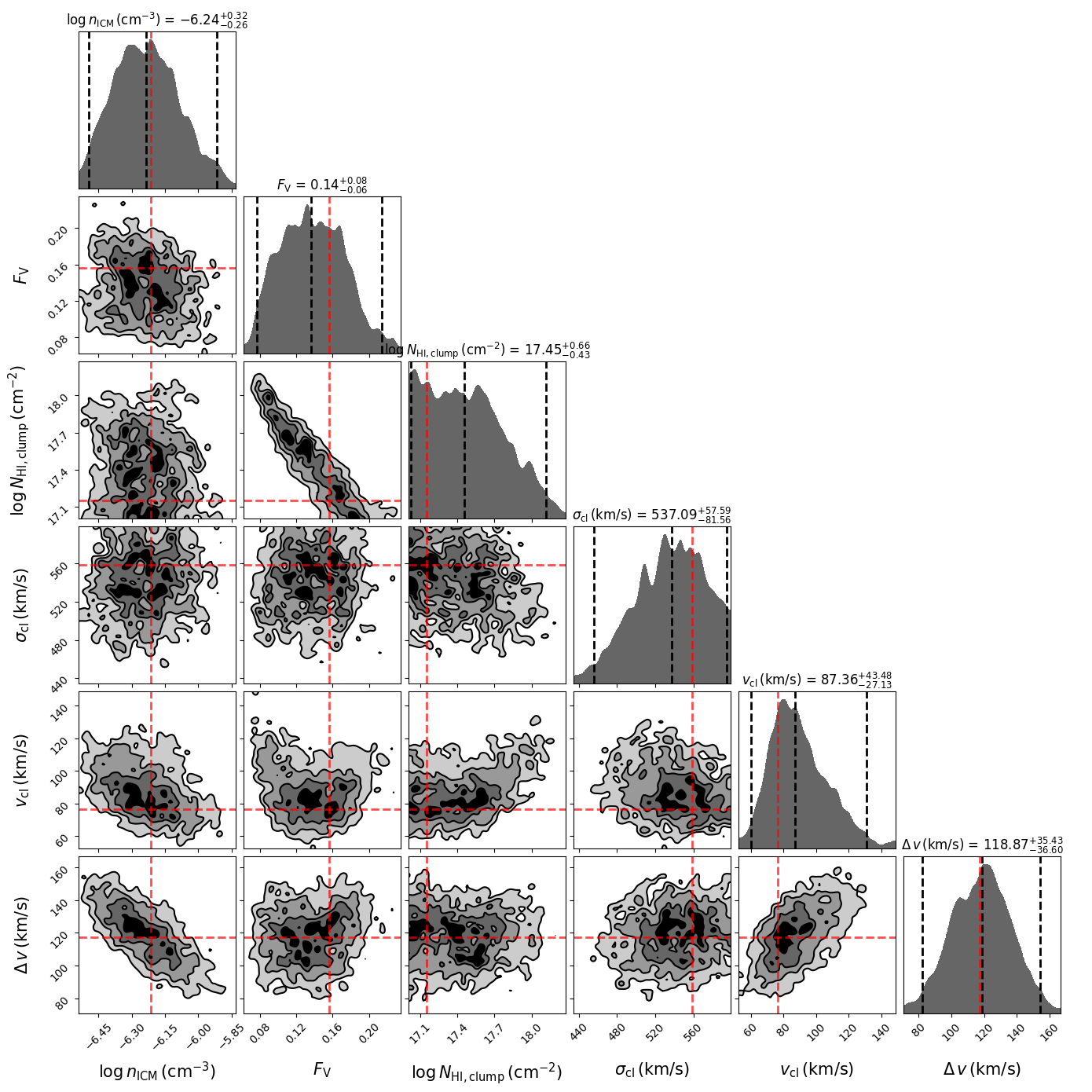}
\end{figure*}

\begin{figure*}
\centering
\includegraphics[width=\textwidth]{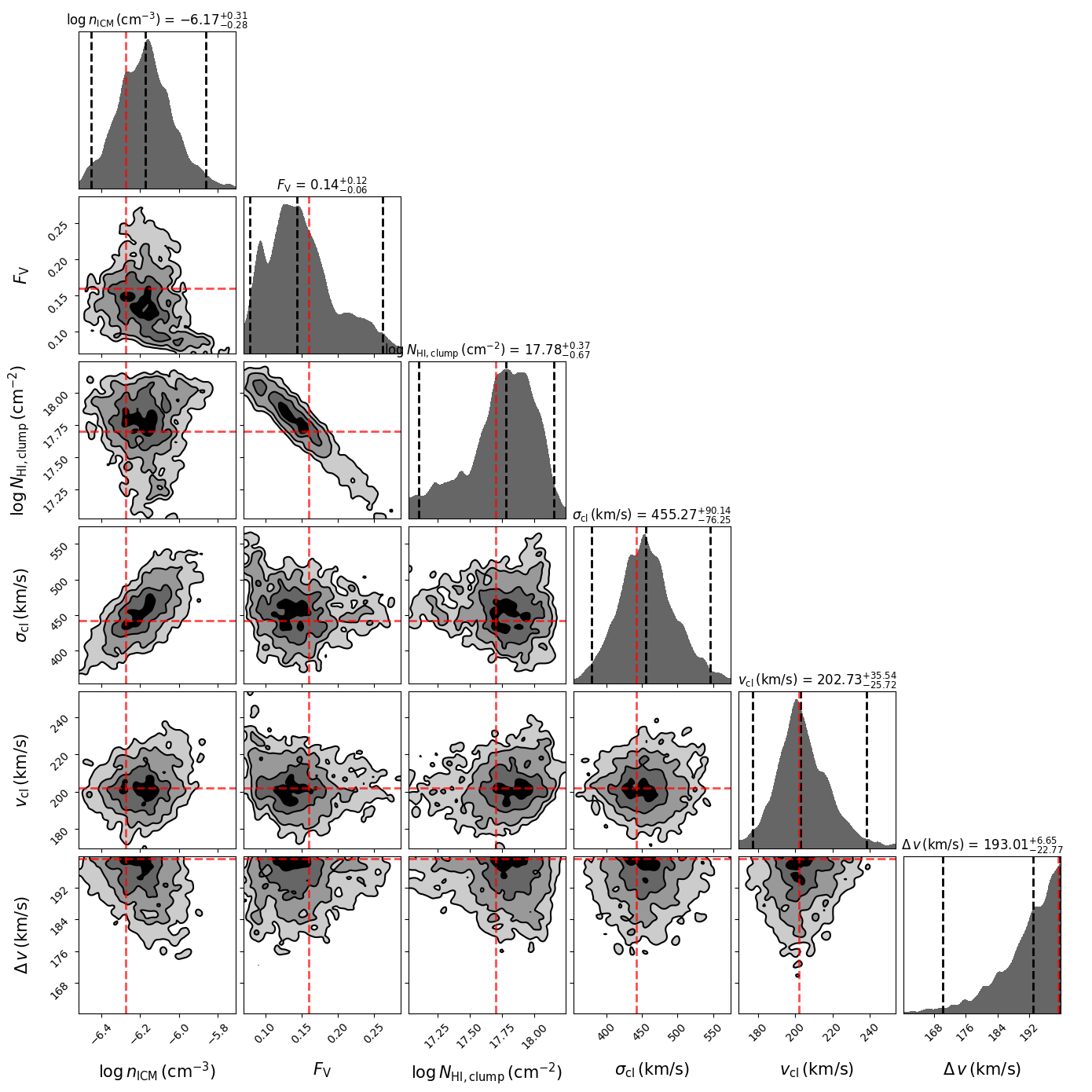}
\end{figure*}

\begin{figure*}
\centering
\includegraphics[width=\textwidth]{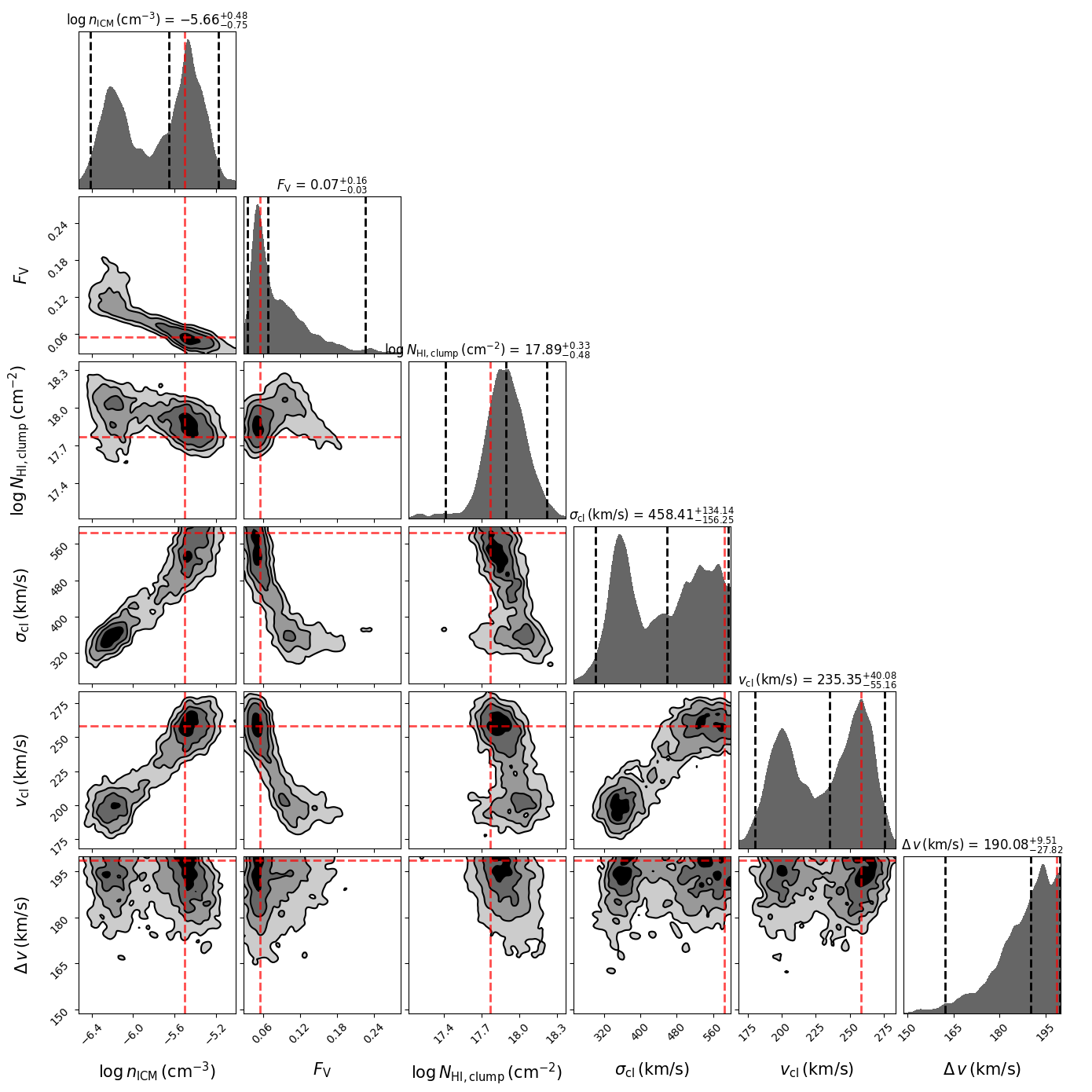}
\end{figure*}

\begin{figure*}
\centering
\includegraphics[width=\textwidth]{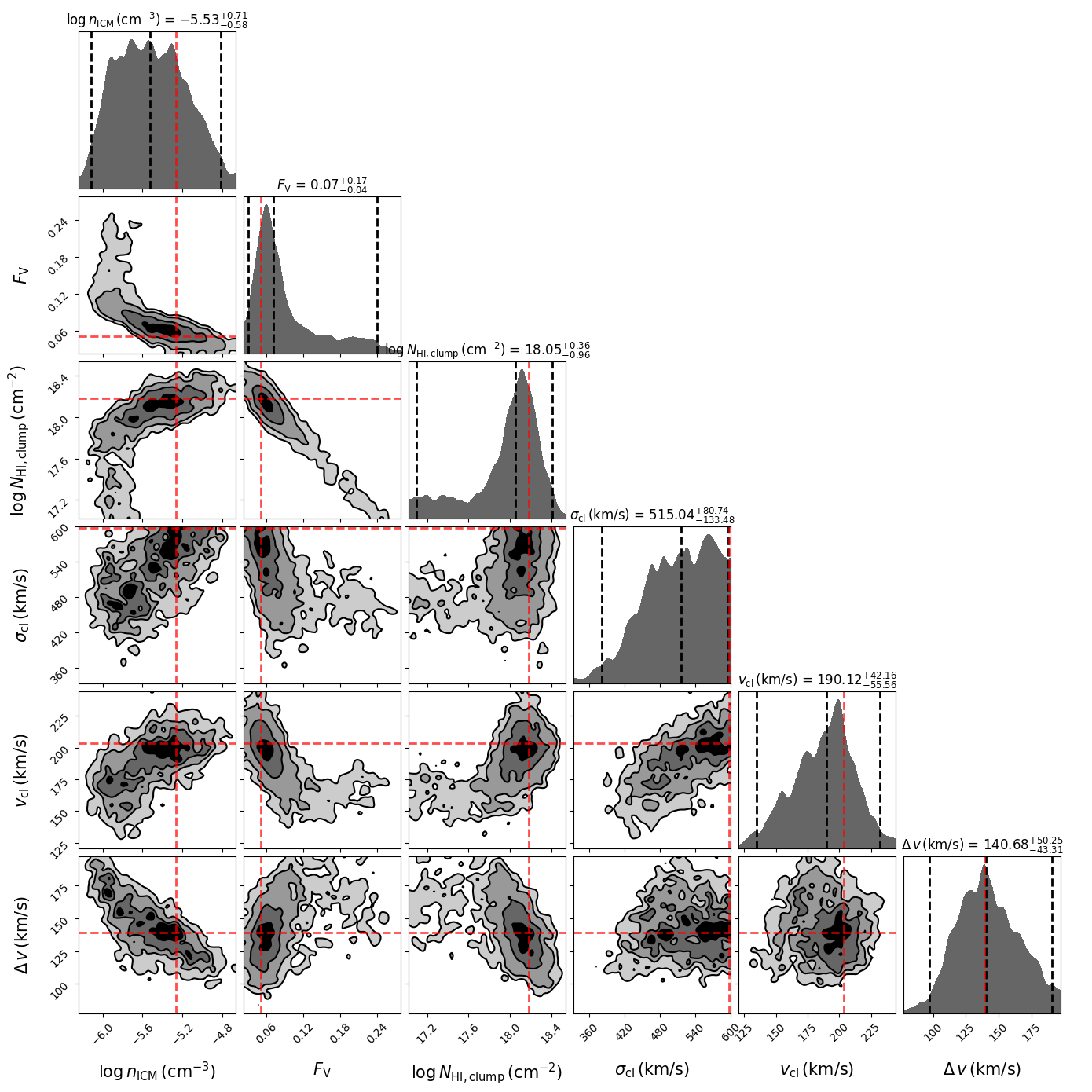}
\end{figure*}

\begin{figure*}
\centering
\includegraphics[width=\textwidth]{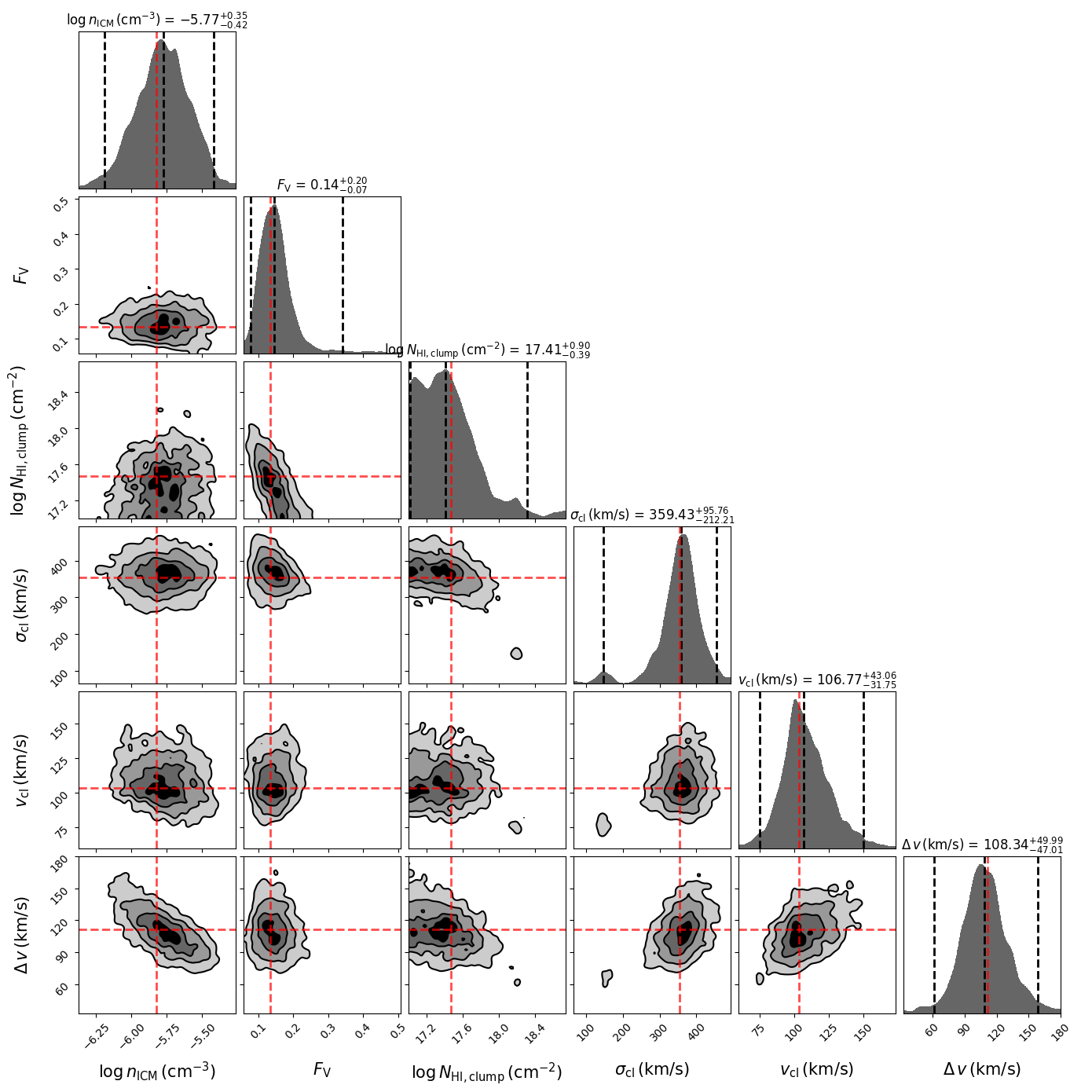}
\end{figure*}

\begin{figure*}
\centering
\includegraphics[width=\textwidth]{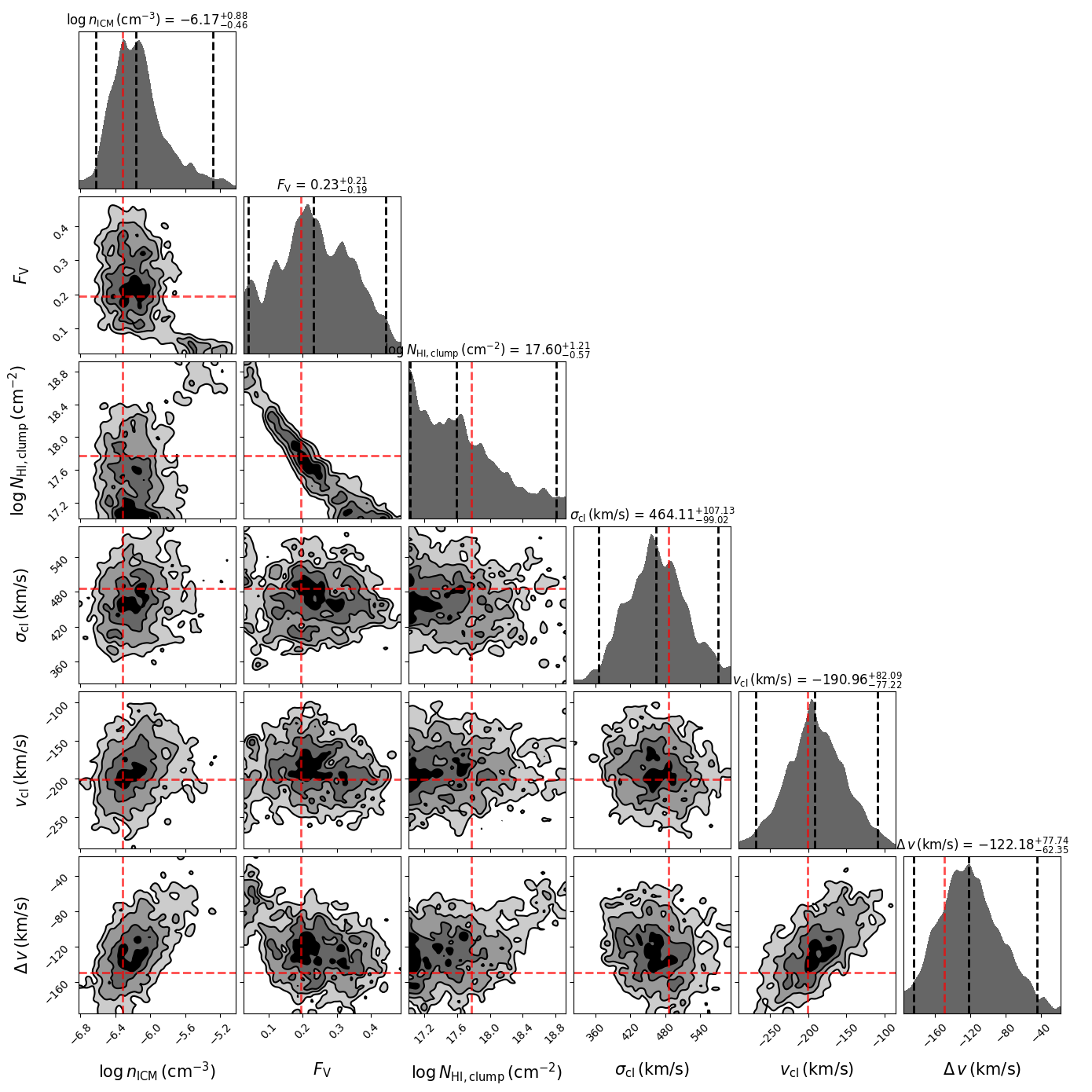}
\end{figure*}

\begin{figure*}
\centering
\includegraphics[width=\textwidth]{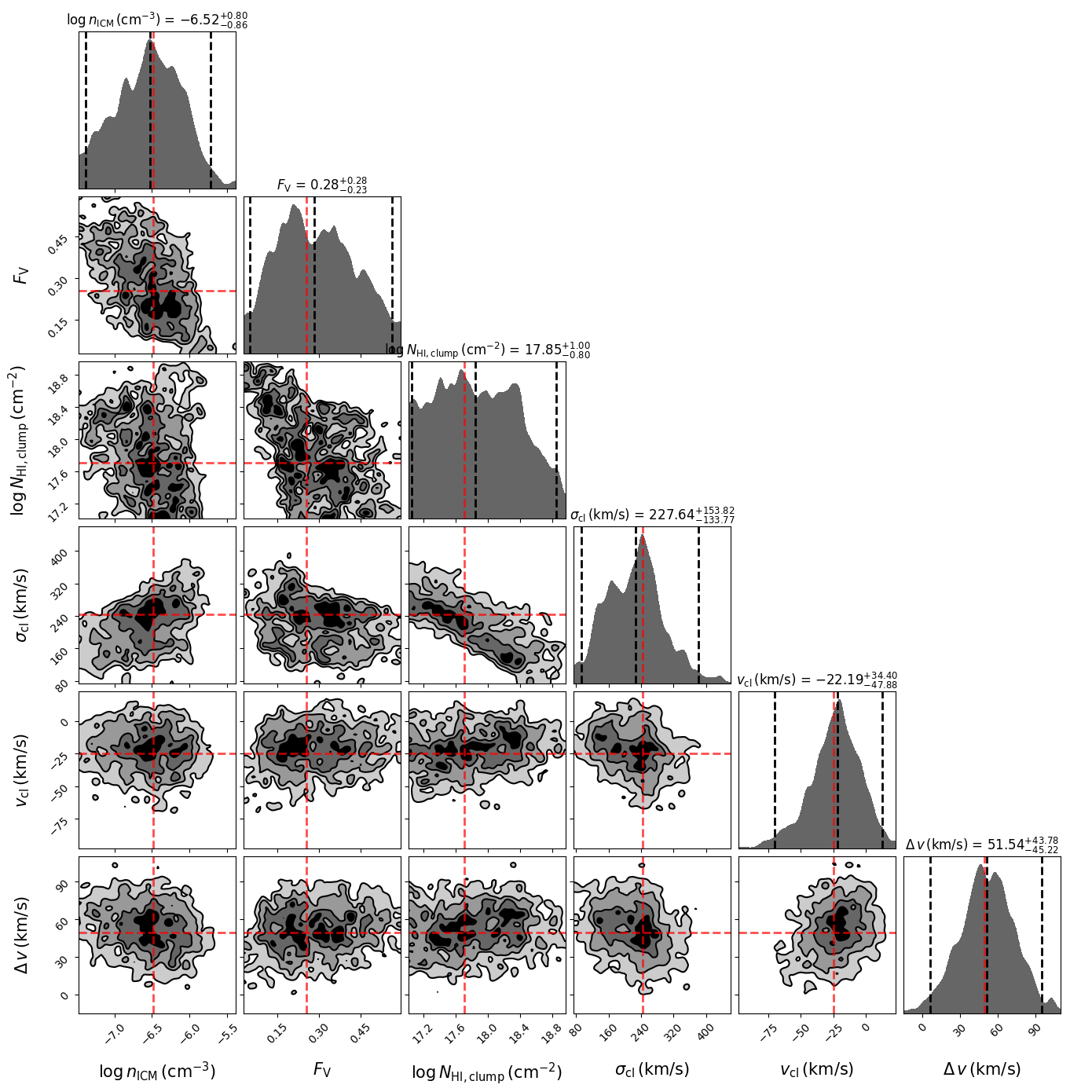}
    \caption{Joint and marginal posterior probability distributions of the multiphase, clumpy model parameters for all fifteen representative \lya\ spectra derived from nested sampling. The vertical black dashed lines indicate the [2.5\%, 50\%, 97.5\%] quantiles (i.e., 2-$\sigma$ confidence intervals). The vertical red dashed lines indicate the locations of the maximum posterior probability. 
    \label{fig:posterior}}
\end{figure*}

\bsp	
\label{lastpage}
\end{document}